\documentclass{PoS}

\usepackage{dcolumn}
\usepackage{graphicx}
\usepackage{bm}
\usepackage{epsfig}

\def\Pi{\mathcal{P}_\infty}

\def\o{\omega}

\leftline{HU-EP-09/54}

\title{Vortex Content of SU(2) Calorons and Multi-Calorons}

\ShortTitle{Vortex Content of SU(2) Calorons}

\author{\speaker{Bo Zhang}\\
	Institut f\"ur Theoretische Physik, Universit\"at Regensburg, D-93040 Regensburg, Germany\\
        E-mail: \email{bo1.zhang@physik.uni-regensburg.de}}

\author{Falk Bruckmann\\
	Institut f\"ur Theoretische Physik, Universit\"at Regensburg, D-93040 Regensburg, Germany\\
        E-mail: \email{falk.bruckmann@physik.uni-regensburg.de}}

\author{Ernst-Michael Ilgenfritz\\
	Institut f\"ur Theoretische Physik, Ruprecht-Karls-Universit\"at Heidelberg, D-69120 Heidelberg, Germany\\
	Institut f\"ur Physik, Humboldt-Universit\"at, D-12489 Berlin, Germany\\
        E-mail: \email{ilgenfri@physik.hu-berlin.de}}

\abstract{We use Laplacian Center Gauge to reveal the vortex content of single 
$SU(2)$ calorons and multi-caloron systems at different holonomies. The vortex 
surfaces
in a single $SU(2)$ caloron 
consist of two parts that are induced by the constituent dyon charges and 
by the twist between the dyons, respectively.
The latter part percolates in a caloron ensemble at maximal nontrivial holonomy. This finding fits perfectly
in the confinement scenario of vortices and shows that calorons are suitable to facilitate the
vortex confinement mechanism.}

\FullConference{The XXVII International Symposium on Lattice Field Theory - LAT2009\\
		 July 26-31 2009\\
		 Peking University, Beijing, China}

\begin{document}

\section{Caloron, Monopole and Vortex}

\enlargethispage{\baselineskip}

	Topological excitations are candidates for the mechanism
for nonperturbative effects in QCD including confinement and chiral condensate.
The most intensively examined topological excitations are instantons, magnetic
monopoles and center vortices. Instantons are 
solutions of the equation of motion, thus
can be introduced into QCD in a semiclassical approach. It can explain
the chiral condensate naturally through the (quasi) zero mode,
%% FB \footnote{This mechanism is based on the index theorem 
%% FB and thus will work for any object with topological charge.}, 
but confinement remained unexplained with instantons. On
the other hand, Abelian monopoles and vortices are not of semiclassical nature,
%% EMI but represent codimension 3 and 2 defects 
but represent defects of codimension 3 and 2, respectively,
that remain after gauge fixings and projections. They can explain confinement, 
and some quantitative studies \cite{Boyko:2006ic} show that they are a prerequisite 
for the occurrence of topological charge in general.

\begin{figure}[b]
\includegraphics[width=0.35\textwidth,bb=-200 220 169 470]{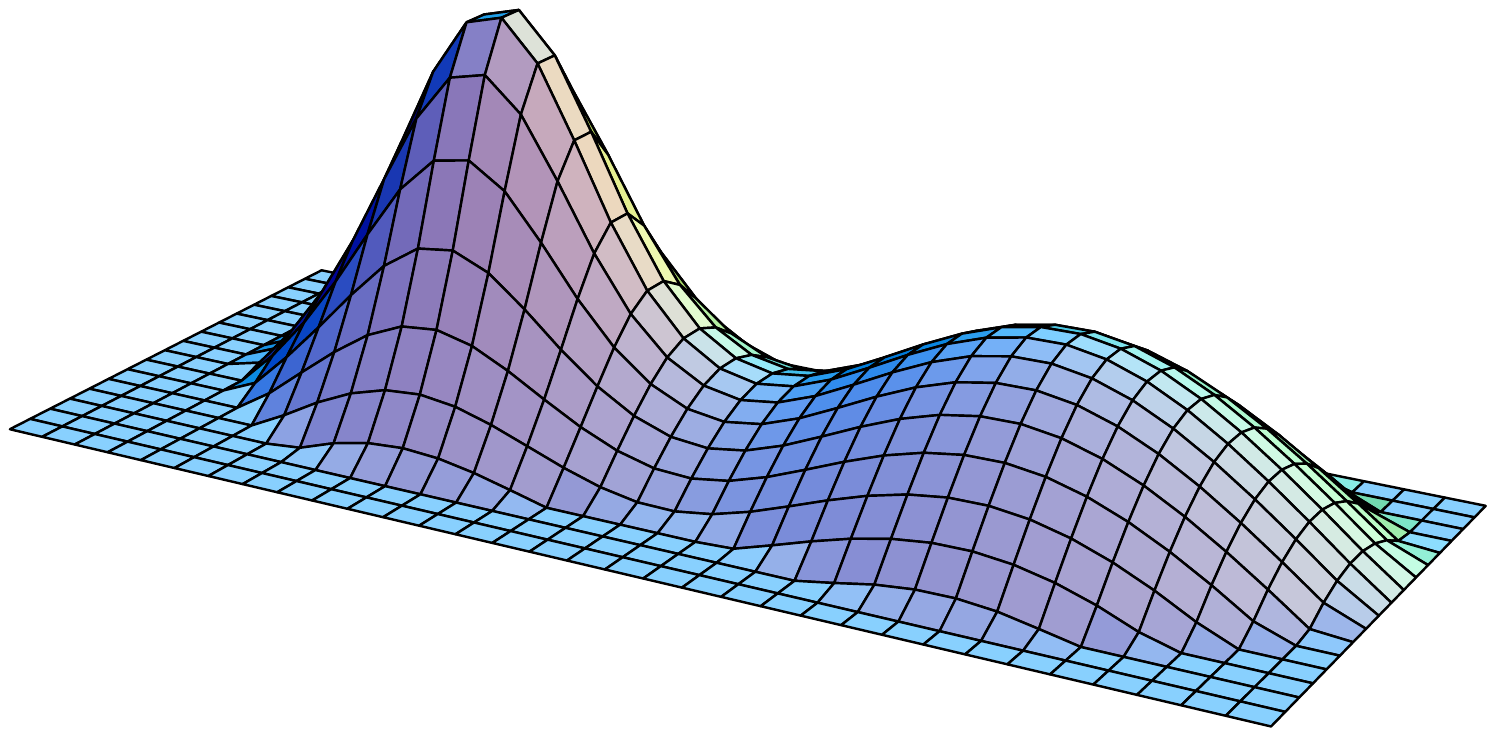}
\vspace{-1.0cm}
\caption{Action density of a $SU(2)$ caloron shown in logarithmic scale, from \cite{Kraan:1998pm}.}
\label{fig:caloron_observables}
\end{figure}

	Calorons \cite{Harrington:1978ve,Kraan:1998pm,Lee:1998bb} 
are generalizations of instantons at finite temperature. The asymptotic Pol\-ya\-kov
loop plays a key role in determining the properties of calorons. In $SU(2)$, we
parameterize it as $P(\vec{x} \rightarrow \infty)=e^{2\pi i \omega \sigma_3}$ with 
the holonomy parameter $\omega$.
A nontrivial holonomy caloron ($\omega\neq0,1/2$) with unit topological charge 
is composed of $N$ dyons (magnetic
monopoles) in gauge group $SU(N)$, namely its action density can have $N$ peaks
located at $N$ constituent dyons 
%% EMI when they 
when these are well separated. A picture for the action
density of a $SU(2)$ charge-one caloron is shown in
Fig.~\ref{fig:caloron_observables}. In $SU(2)$, a charge-one caloron has two
constituent dyons, one has magnetic charge +1 (M dyon), another has magnetic 
charge -1 (L dyon), the
distance between the two dyons is $\pi \rho^2 /\beta$, where $\rho$ is the size 
parameter of the caloron
%% FB 
(and $\beta=1/k_BT$).
%% FB
Furthermore, 
%% EMI these two dyons have a relative unit 
%% EMI twist in time direction 
one of the dyons has a 
%% FB relative, time dependent 
twist of unit one relative to 
the other dyon
%% EMI (a twist is a nontrivial gauge transformation in time direction here). 
(here, a twist is a nontrivial gauge transformation in time direction). 
In periodic gauge, the M dyon is approximately static
while the L dyon 
%% EMI has 
carries a unit twist \cite{Kraan:1998pm}. It can be shown
that the topological charge of the caloron equals the magnetic charge times 
the relative twist of its constituent dyons \cite{Kraan:1998pm}. In this way, 
calorons and magnetic monopoles
%% EMI are somehow 'unified'.
are close relatives.

We will think of the traced holonomy ${\rm tr}\, P(\vec{x} \rightarrow \infty)=2\cos(2\pi\omega)$ as identified with the center symmetry order parameter $\langle {\rm tr}\, P \rangle$, such that maximal nontrivial holonomy $\omega=1/4$ stands for the confined phase, whereas other holonomies amount to temperatures $T>T_c$.

	Vortices and magnetic monopoles are also closely
related with each other. In four dimensional space time, vortices form two
dimensional world sheets while magnetic monopoles are one dimensional world
lines. In the combination of Laplacian Abelian Gauge (LAG \cite{vanderSijs}) and 
Laplacian Center Gauge (LCG \cite{deForcrand:2000pg}), magnetic
monopole worldlines reside on the vortex sheets \cite{Alexandrou:1999iy}.

	Here we are going to find the relation between $SU(2)$ caloron and
vortices. This includes the vortices in individual calorons depending on
the holonomy, the intersection of different parts of vortices recovering the topological charge and
the dependence of vortices in a caloron gas on the holonomy which 
%% FB have 
has an interesting 
relation to percolation and confinement. 
%the vortices in a caloron gas depending on the holonomy which results in an interesting relation to 
%percolation and confinement. 

\section{Laplacian Center Gauge}

There are different gauge fixing methods to find vortices, such as Direct
Maximal Center Gauge (DMCG) \cite{DCG} and Indirect Maximal Center Gauge (IMCG)
\cite{IMCG}. 
%% FB In both gauges, one 
%% EMI minimizes .... really ?
%% FB maximizes the non-center part of the links
%% EMI actually the trace squared or the adjoint trace
%% FB (namely the center-blind adjoint links) at the expense of a Gribov copy problem.
An alternative is the Laplacian Center Gauge (LCG), which can be 
viewed as a generalization of DMCG avoiding the Gribov copy problem \cite{deForcrand:2000pg}.
To this end one computes the two lowest eigenvectors ($\phi_0,\phi_1$) of the gauge covariant 
Laplace operator in the adjoint representation, which we do by virtue of 
ARPACK.
%% FB the ARPACK package.
%\begin{eqnarray}
% -\Delta [U^A]\phi_{0,1}& = & E_{0,1}\phi_{0,1}
%\label{eqn_eigen}\\
% \Delta^{ab}_{xy} [U^A] & = & \frac{1}{a^2}\sum_\mu 
% \Big(   U_\mu^A(x)^{ab} \delta_{x+a\hat\mu,y} 
%\Big.    \Big. 
%+ U_\mu^A(x-\hat{\mu})^{ba} \delta_{x-a\hat{\mu},y}  - 2 \delta^{ab} \delta_{xy}
%\Big), 
%\end{eqnarray}

	In LCG, the first step is to rotate the lowest mode $\phi_0$ to the third color
direction with a gauge transformation $V$, 
%% FB i.e.\ diagonalised, 
$^V\!\phi_0=|\phi_0|\,\sigma_3\ $,
i.e.\ to diagonalize it. 
The remaining
Abelian freedom of rotations around the third axis, 
$V \to vV$ with $v=\exp(i\alpha\sigma_3)$, is fixed (up to center 
elements) by demanding a particular form of
the first excited mode,
%% FB the  $^{vV}\!\phi_1$ with vanishing second component and positive first component respectively:
 $(^{vV}\!\phi_1)^{a=2}=0$ and $(^{vV}\!\phi_1)^{a=1}>0 $.

	Defects of this gauge fixing appear when $\phi_0$ and $\phi_1$ 
are collinear, because then the Abelian freedom parameterized by $v$ remains 
unfixed. It was shown in \cite{deForcrand:2000pg} that points $x$
where $\phi_0(x)$ and $\phi_1(x)$ are collinear define the (generically 
two-dimensional) vortex surface. This includes points $x$, where $\phi_0$ 
vanishes, $\phi_0(x)=0$, which define monopole worldlines in the 
Laplacian Abelian Gauge (LAG) \cite{vanderSijs}.

\begin{figure}[b]
\begin{center}
\includegraphics[width=0.57\linewidth,bb=-10 -10 300 130]{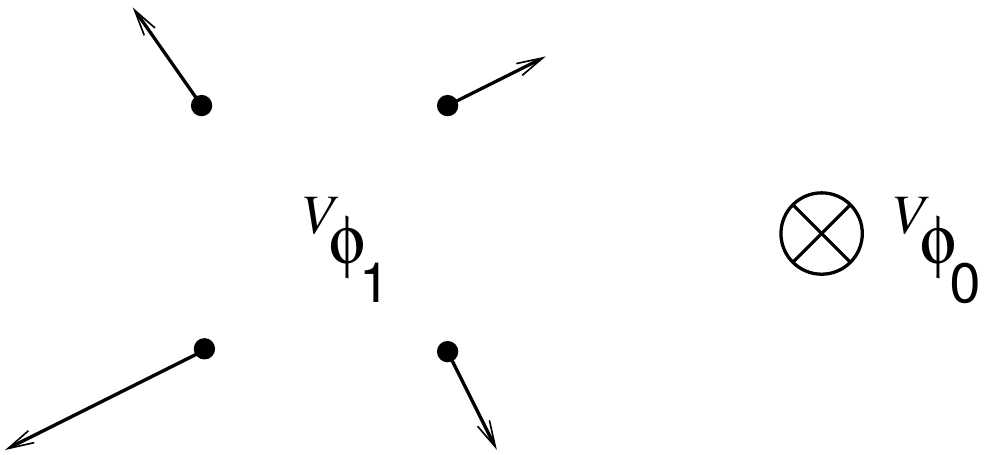}
\vspace{-0.4cm}
\end{center}
\caption{The topological argument to detect vortices on a plaquette: 
Components of the first excited mode $\phi_1$ that are perpendicular 
to the lowest mode $\phi_0$ (after both have been gauge transformed 
by $V$) are plotted on all sites of the plaquette. The configuration 
shown here has a nontrivial winding number, which implies that the 
two eigenvectors are collinear in color space somewhere inside the 
plaquette.}
\label{fig:winding}
\end{figure}

We detect the center vortices in LCG with the help of a topological 
argument: after having diagonalised $\phi_0$ with $V$, the question whether $\phi_0$ and $\phi_1$
are collinear amounts to $^V\!\phi_1$ being diagonal too, i.e. having zero
nondiagonal components. We therefore inspect the projection of 
$^V\!\phi_1$ onto the $(\sigma_1,\sigma_2)$-plane for all four points of 
each plaquette, see Fig. \ref{fig:winding}.
By assuming continuity of the field 
%% FB (more precisely, its $(\sigma_1,\sigma_2)$-projection) 
on 
%% FB a 
the plaquette,
a nontrivial winding number of the 2d vector $(^V \phi_1^1, ^V\phi_1^2) $
around the edge of the plaquette 
%% EMI means 
indicates 
%% EMI indicates the presence of 
a zero inside. In this case
we say that the midpoint of that plaquette belongs to the vortex surfaces.
%% EMI make a full sentence out of the following:
The vortex sheet extends in dual directions from the midpoint. It
consists of plaquettes of the dual lattice shifted by $a/2$ in all directions 
with respect to the original lattice.
%% FB  (see \cite{Caloron and Vortex} for details).

%The problem of such a process is that the gauge transformation that fix $\phi_0$ to the north pole as
%a function of $\phi_0$ is a mapping from $ \vec{\phi_0} \sim
%S^2$ to $S^3 \sim SU(2)$. A simplest non-singular of such a mapping
%should have a unit winding number, but $\pi_2(S^3)=0$ tells us
%that singular points always exist in such a mapping, this point will ruin the
%continuity assumption. For example, if we define such a mapping as
%\ref{eqn_gauge_fixing_transformation}, the singular points
%is the south pole. When the lowest eigenvector $\phi_0$ is in the
%negative $\sigma_3$-direction, the corresponding diagonalising gauge
%transformation $V$ changes it drastically (like when ``combing a hedgehog''). In this case,
%the correspondingly transformed first excited state, $^V\!\phi_1$ also changes drastically and the
%continuity assumption can result in artificial winding numbers and thus
%unphysical vortices.

%\begin{eqnarray}
%V=&&\frac{1}{Y}
%  \begin{pmatrix}
%    |\phi_0|+\phi_0^3              &   \phi_0^1 -i \phi_0^2  \\
%    -\phi_0^1 -i \phi_0^2          &   |\phi_0|+\phi_0^3  \\
%  \end{pmatrix}  \\
%Y=&&\sqrt{    (\phi_0^1)^2   + ( \phi_0^2)^2   +
%(|\phi_0|+\phi_1^3 )^2}
%\label{eqn_gauge_fixing_transformation}
%\end{eqnarray}

The first step of LCG, i.e. $^V\!\phi_0=|\phi_0|\,\sigma_3\ $ can include gauge transformations 
varying rapidly in space, this would result in artificial winding numbers and 
thus unphysical vortices. However, to detect vortices, the lowest eigenvector can be fixed 
to any direction \cite{deForcrand:2000pg}, i.e.\ fixed to different
%% FB
color
%% FB
directions on different points. Thus, plaquette by plaquette we rotate $\phi_0$ to 
its average direction over the four sites on the plaquette (which in most cases 
is a small rotation), afterwards we inspect $\phi_1$ in the plane perpendicular 
to that direction.

%Note that the winding number changes sign under $\phi_0\to -\phi_0$, but not
%under $\phi_1 \to -\phi_1$. Both sign changes do not change the fact that these fields
%are eigenmodes of the Laplacian and hence the global signs of $\phi_0$, $\phi_1$
%and also absolute signs of the winding numbers are ambiguous (but the relative sign of 
%the winding number can be determined, that is the reason we can determine the topological
%charge of the resulting vortices). 

\section{Vortex Surface in a Single Caloron}

To reveal the vortex surface in unit charge caloron, we discretise it on $N_0=8$
and space extension $N_i\gg8$ lattices with constituent dyons 
%% FB fixed 
on the $x_3$-axis and the center of mass fixed at the origin. After that, we compute
the two lowest eigenvectors of the Laplacian operator in the adjoint representation. 
The lowest mode we get does not depend much on the size 
of the lattice, but the first excited state depends on the ratio of 
$N_3$ to $N_{1,2}$. We find that when $N_3/N_{1,2}$ is larger than 1 (such as $80/48$), 
the first excited mode is a singlet, when this
ratio is equal to or smaller than 1 (such as $64/64$), the first excited mode is
a doublet.

This ambiguity reflects the fact that we are forcing states of a continuous 
spectrum into a finite volume, which is similar to waves in a potential well thus
sensitive to boundary conditions (such as $N_i$). Gross features of the adjoint spectrum 
in a caloron background can be understood from the spectrum in a constant link background
with the same holonomy, see \cite{Caloron and Vortex} for details.

%Localized lowest mode, on the contrary,
%should not depend much on the discretization (for instantons compactified 
%on the sphere, level crossings of the adjoint modes as a function of the 
%compactification radius have been observed in Fig.~1 of \cite{Bruckmann:2000ay}).

\begin{figure}[b]
\includegraphics[width=0.38\linewidth,bb=-30 -80 339 170]{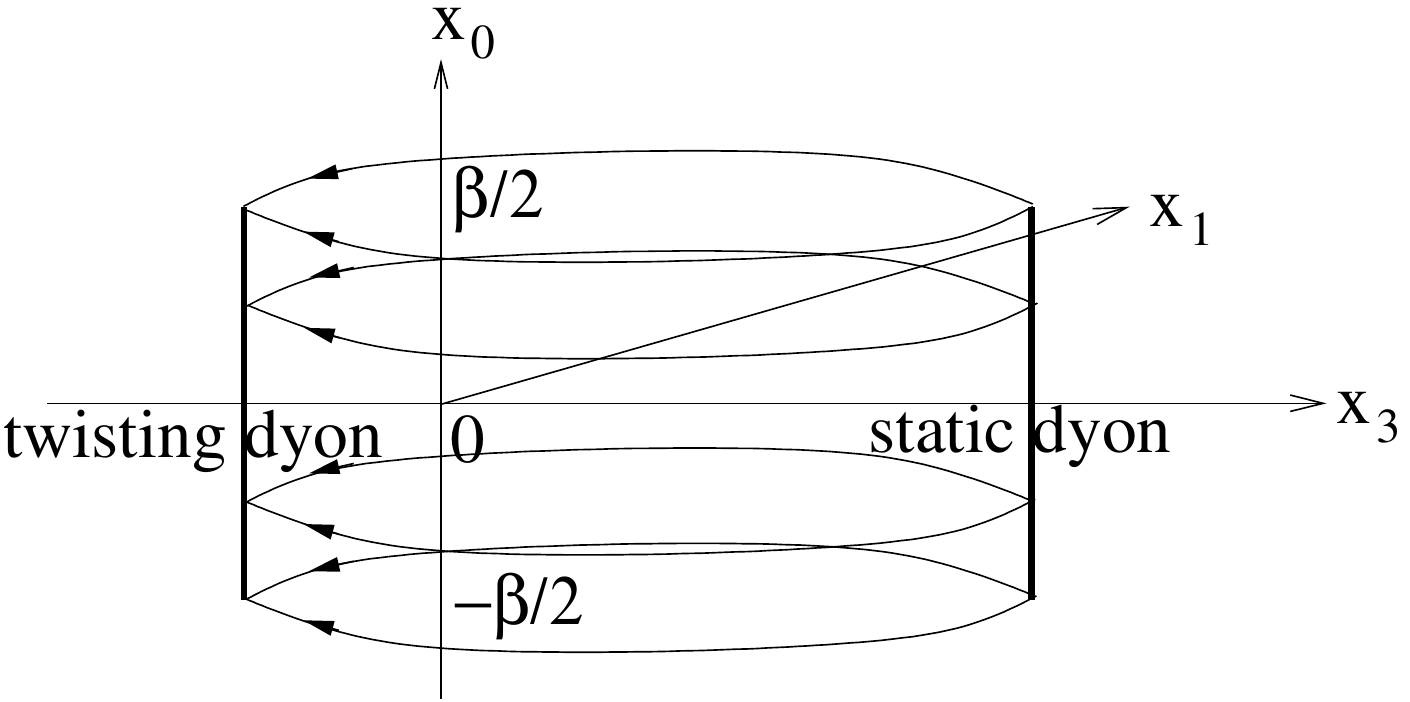}
\hspace{1.5cm}
\includegraphics[width=0.38\linewidth,bb=-30 -80 339 170]{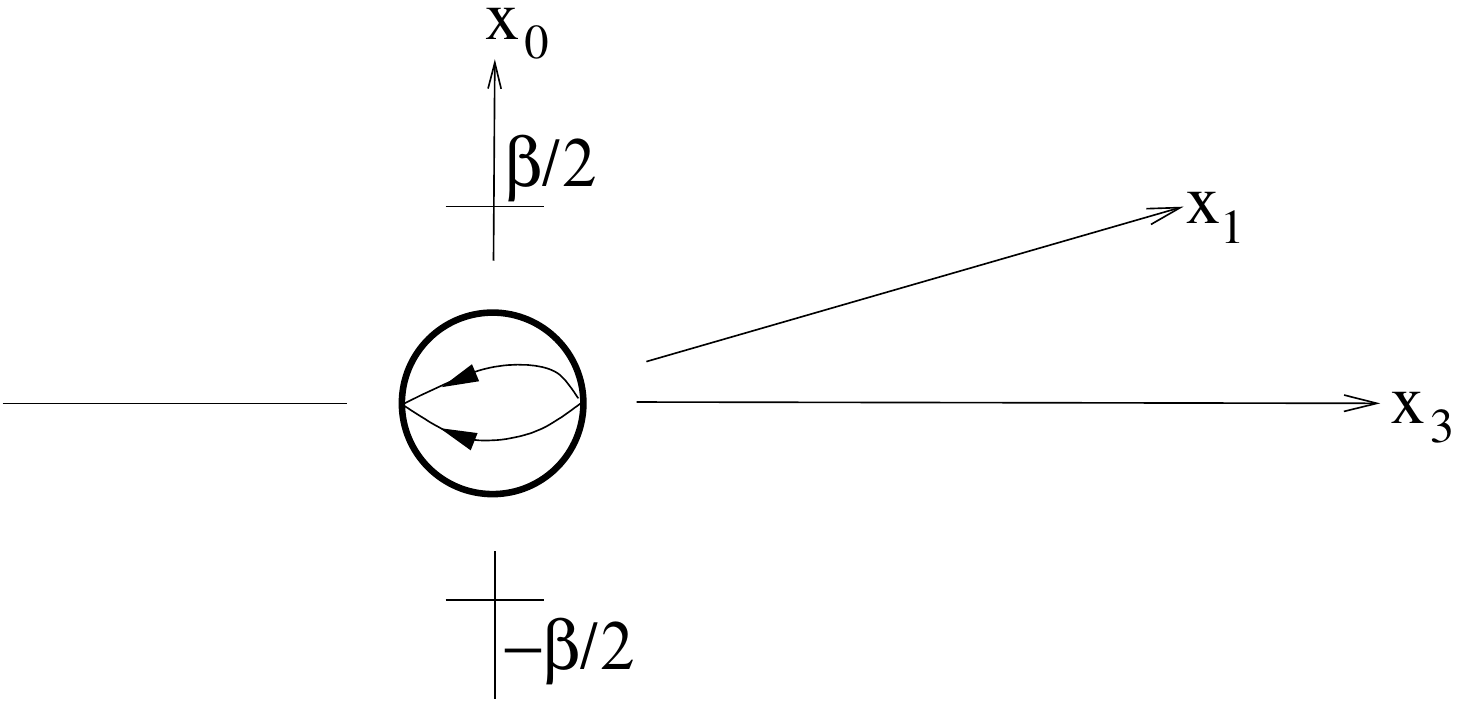}\\
\includegraphics[width=0.38\linewidth,bb=-30 -80 339 170]{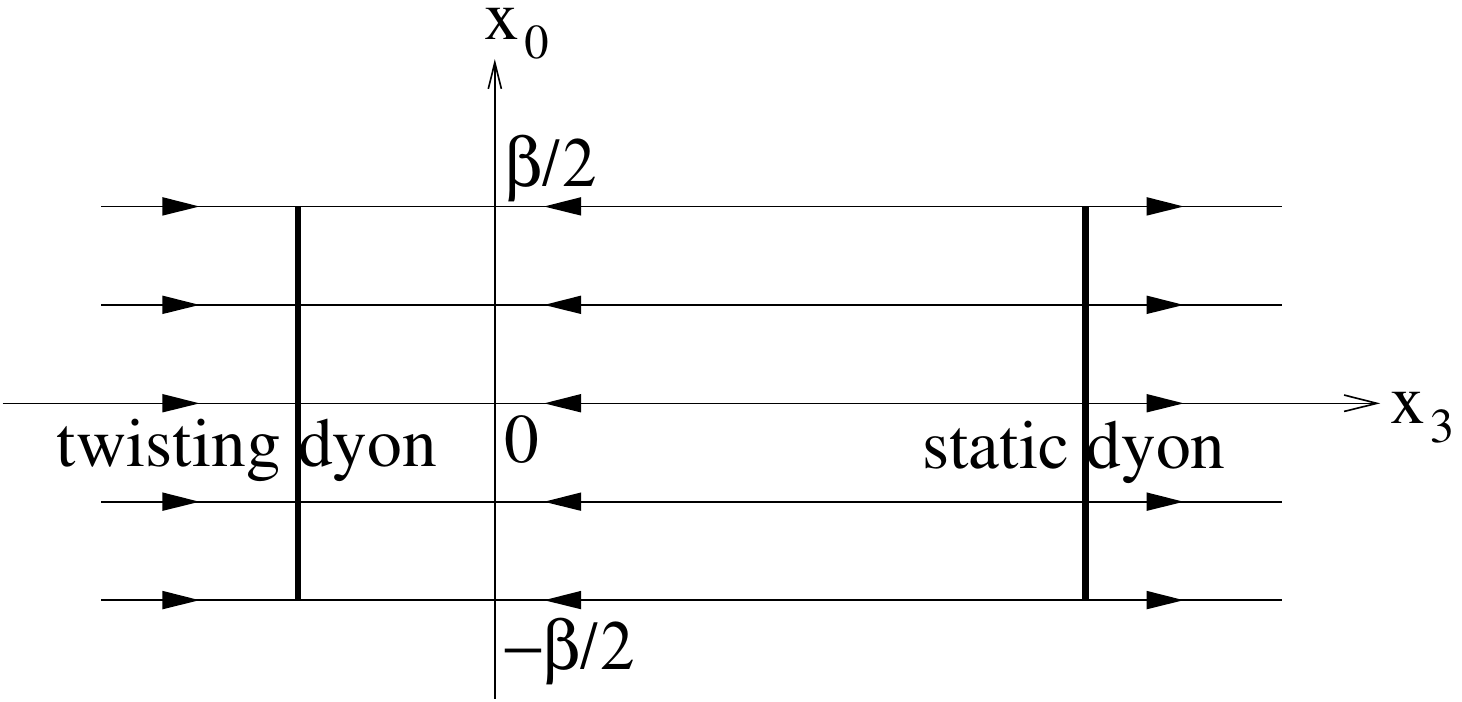}
\hspace{2.0cm}
\includegraphics[width=0.38\linewidth,bb=-30 -80 339 170]{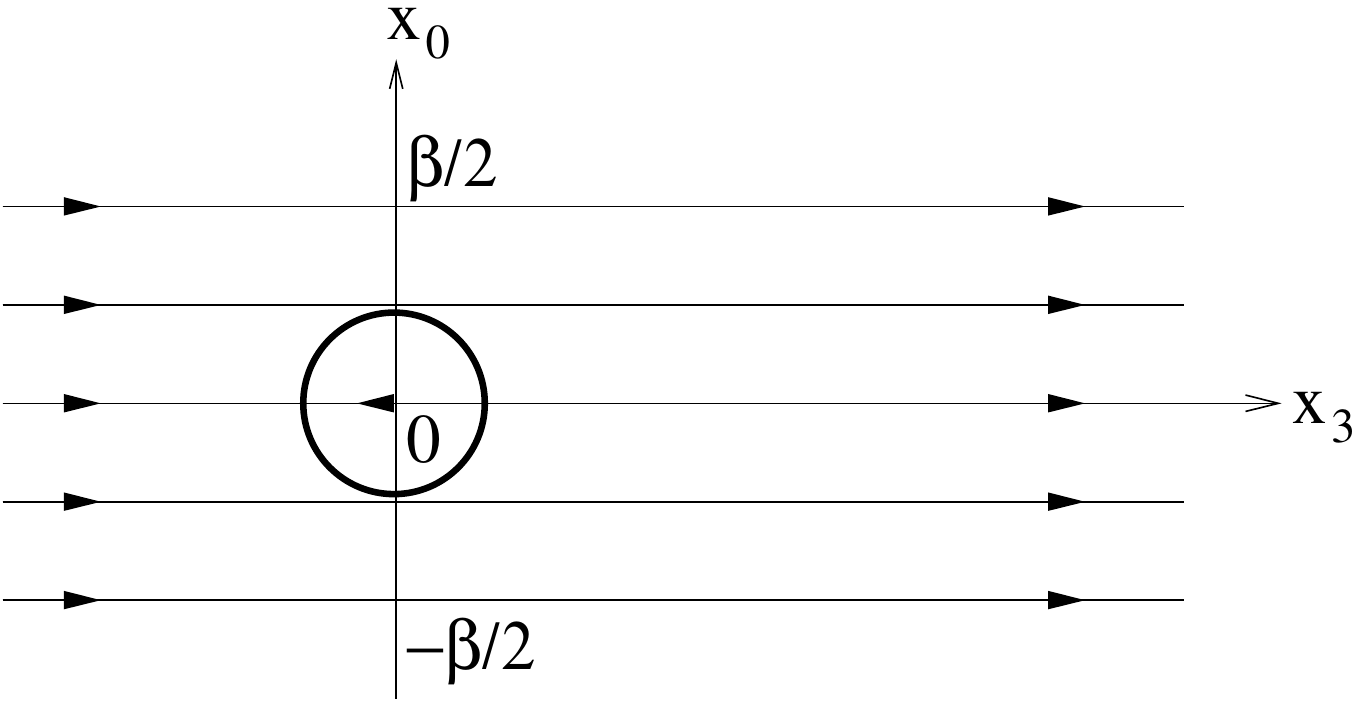}\\
\vspace{-1.9cm}
\caption{Schematic pictures for the dyon charge induced vortex part. 
The upper row is obtained if 
%% EMI for that from the singlet 
the first excited mode is part of a doublet. 
The vortex forms a tube or sphere.
The lower row 
%% EMII row for doublet ones 
corresponds to the case of the first excited mode being a singlet.
The vortex is spanning the $(x_3,x_0)$-plane.
%% EMI while in both rows
The cases of large caloron size parameters $\rho$ are shown on the left, 
the small $\rho$ cases (with closed monopole worldlines in the LAG) on the right. }
\label{dyon charge induced V.S}
\end{figure}

Using the 'local' LCG we mentioned in the last section, we find that the 
%% EMI vortices
vortex surface 
in a unit charge caloron 
consists of two parts, a dyon charge induced part
(shown in Fig.~\ref{dyon charge induced V.S}) and a twist induced part.

%% EMI The dyon charge induced part from doublet first excited modes 
%% EMI is composed of two lines 
If the first excited mode belongs to a doublet,
there are two vortex lines connecting the M-dyon with the L-dyon
%% EMI from M-dyon to L-dyon 
in every time slice where they exist. For
large $\rho$ caloron ($\rho\gtrsim0.5\beta$), the M-dyon and L-dyon exist in all
time slices. Then the vortex sheet
%% EMI and this part 
is completely in space-time direction. For a small $\rho$ caloron, the monopole 
world line is closed and extends over a finite time interval.
%% EMI of the time slices. 
The vortex sheet forms a 
%% FB bubble 
sphere
that includes also space-space plaquettes.
%% EMI direction plaquettes).
%% FB If we take 
Taking another eigenvector in the doublet, the vortex lines will be
rotated 
%% FB by 90 degrees 
around the $x_3$ axis (in all time slices). This reflects 
the axial symmetry of the caloron. 

If the first excited mode is a singlet,
%% FB
on the other hand,
%% FB
%% EMI The dyon charge induced vortices from the singlet first excited mode 
%% EMI is the $x_3$ axis in all
%% EMI time slices. 
the dyon charge induced vortex sheet is the $(x_3,x_0)$-plane which contains the
monopole world lines.
%% ZB  In each time slice, the flux on it changes its sign at the dyons.
%% ZB  I cut 'In each time slice' because dyons does not exist in each time slice if we have a small $rho$
The flux on it changes its sign at the dyons.
%% EMI means 
Also in this case there are two fluxes running from one of the dyons to the other.\\
%% EMI another.

\begin{figure}[b]
\vspace{-3.0cm}
\includegraphics[width=0.32\linewidth,bb=0 220 369 470]{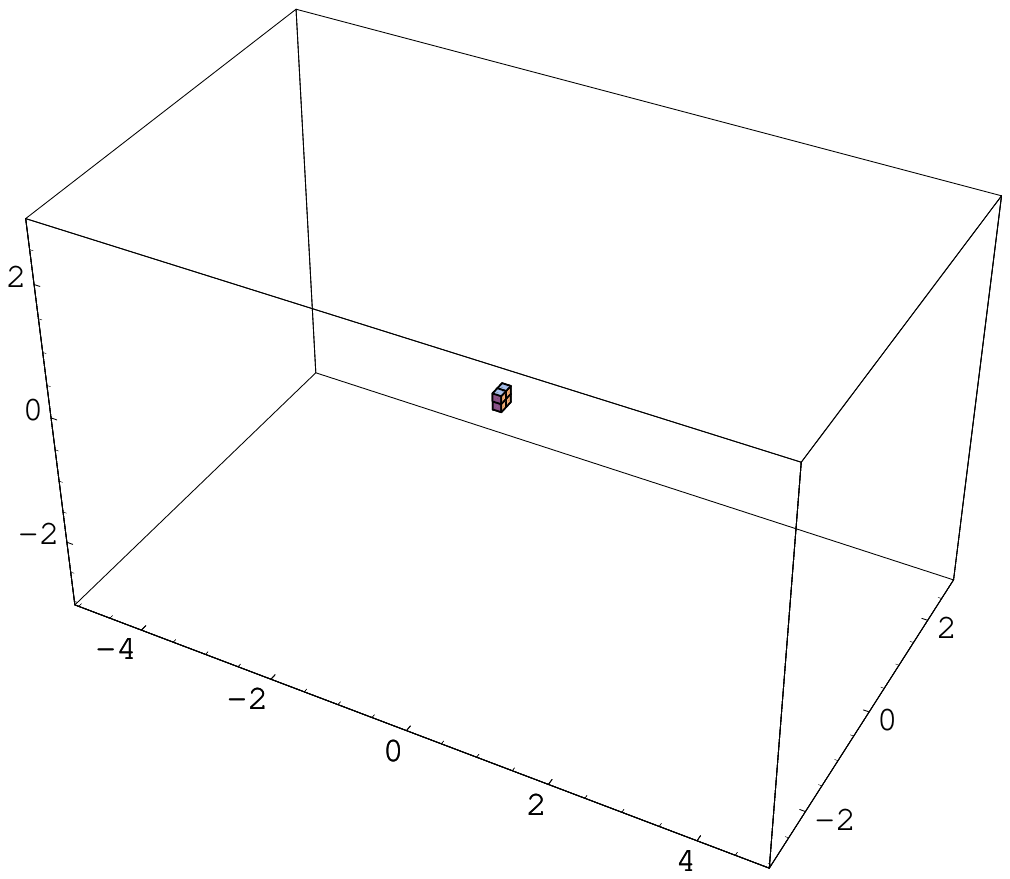}
\includegraphics[width=0.32\linewidth,bb=0 220 369 470]{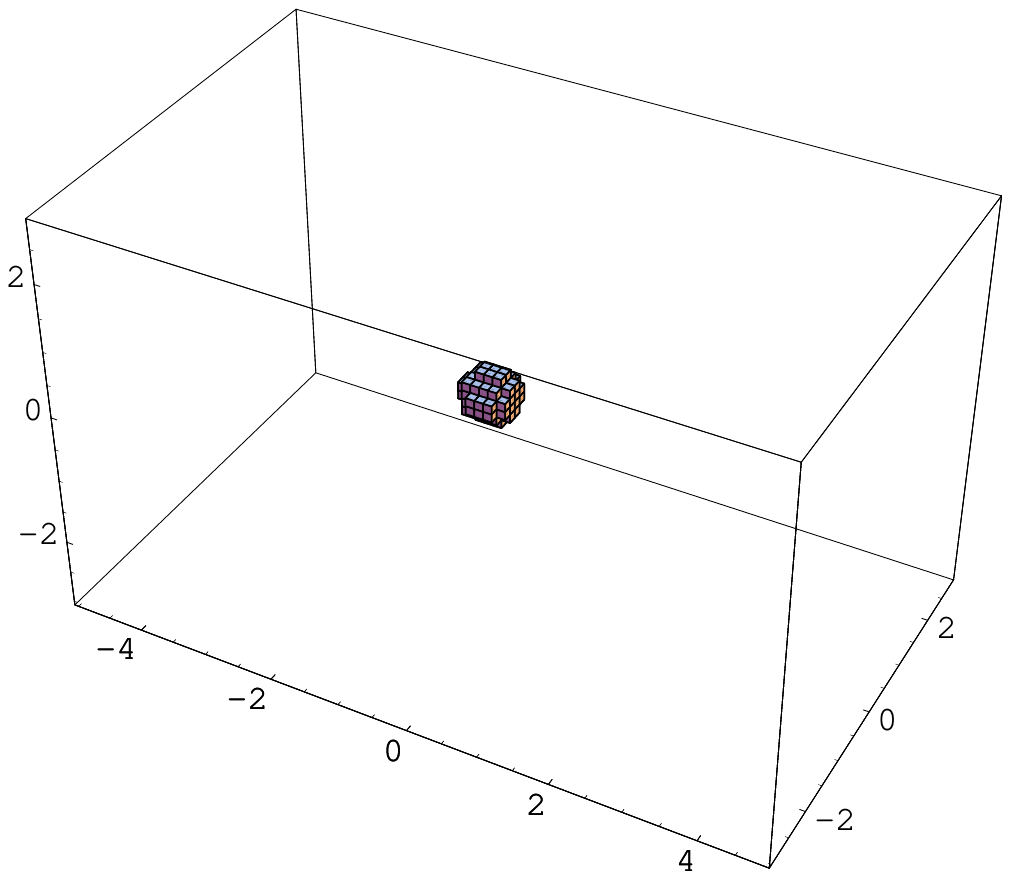}
\includegraphics[width=0.32\linewidth,bb=0 220 369 470]{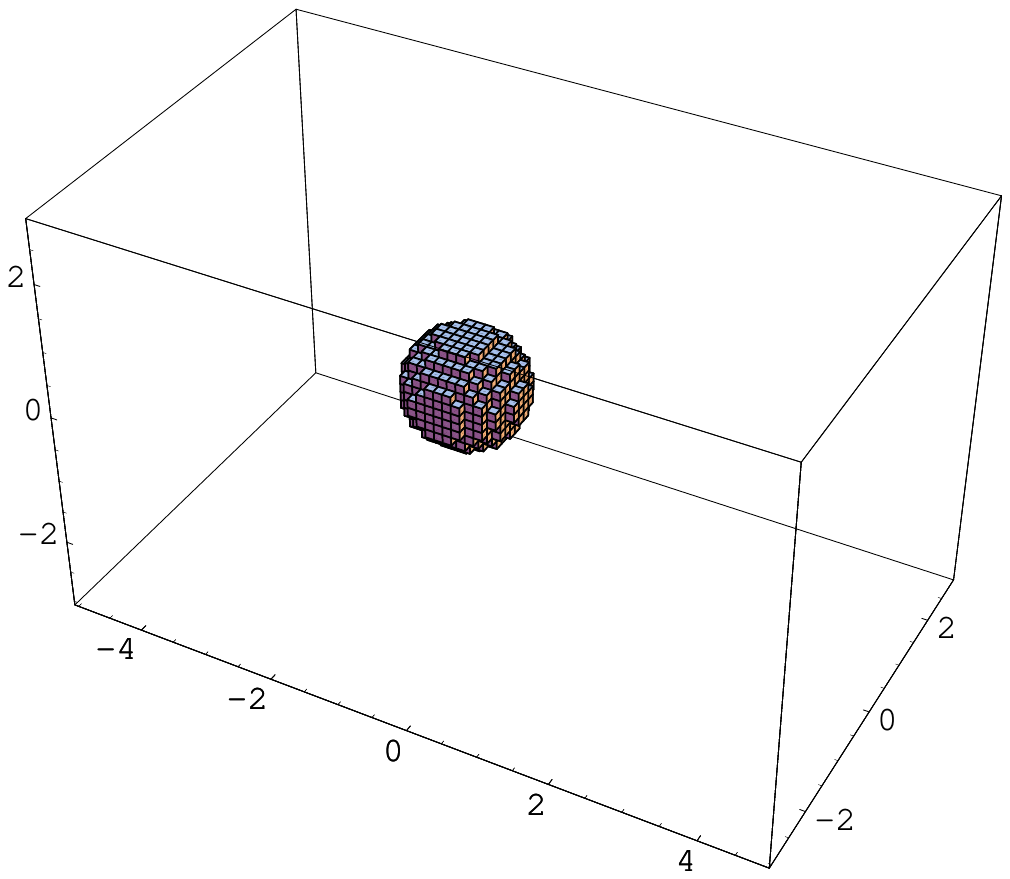}\\
%\vspace{1cm}
\includegraphics[width=0.32\linewidth,bb=0 220 369 470]{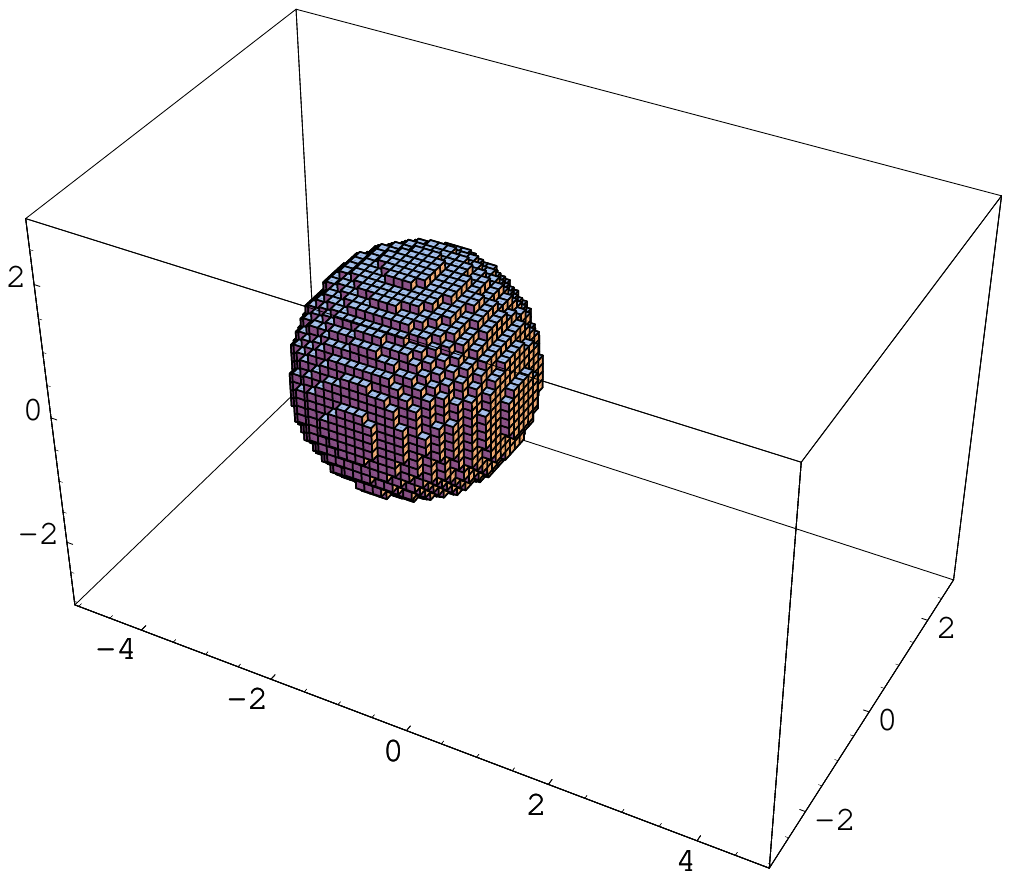}
\includegraphics[width=0.32\linewidth,bb=0 220 369 470]{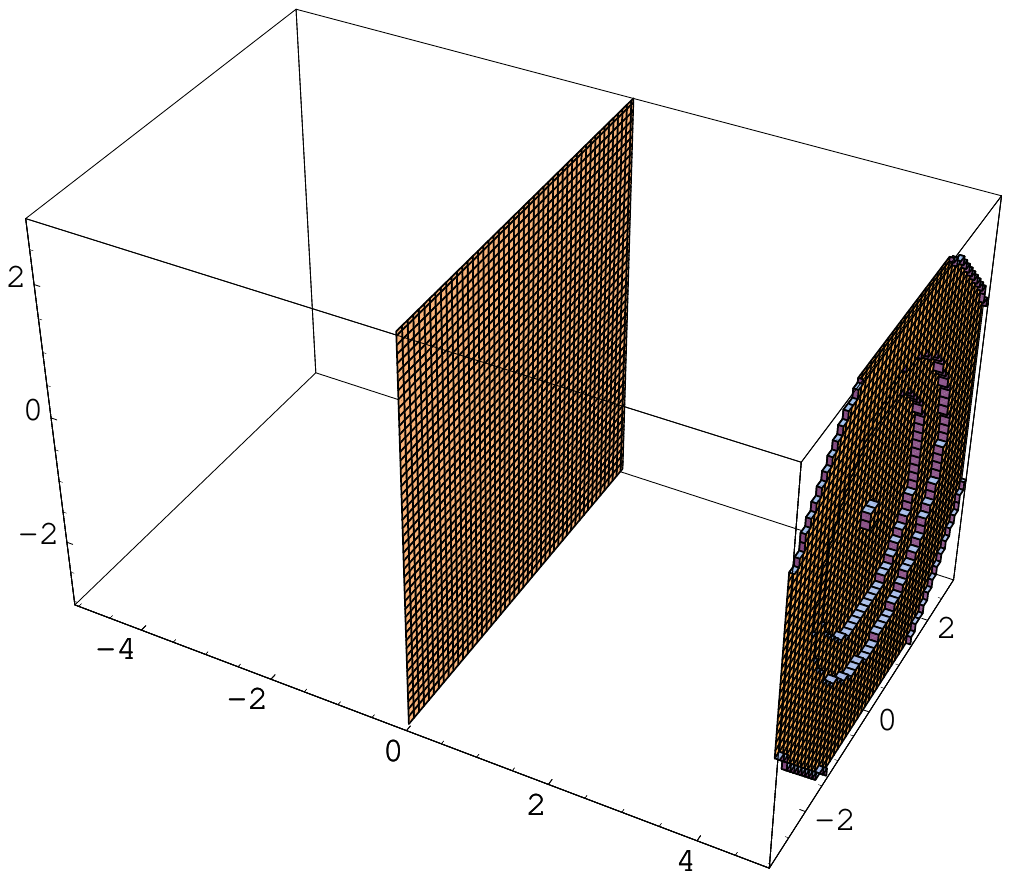}
\includegraphics[width=0.32\linewidth,bb=0 220 369 470]{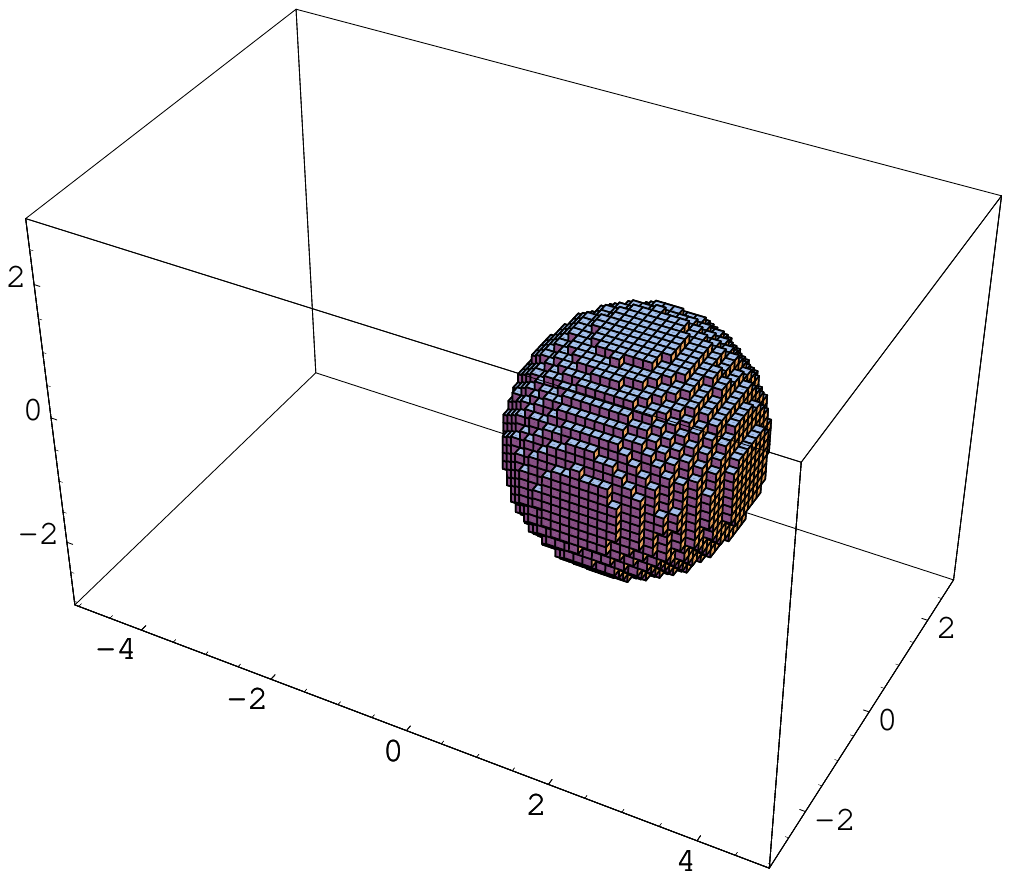}\\
\vspace{0.4cm}
\includegraphics[width=0.32\linewidth,bb=0 220 369 470]{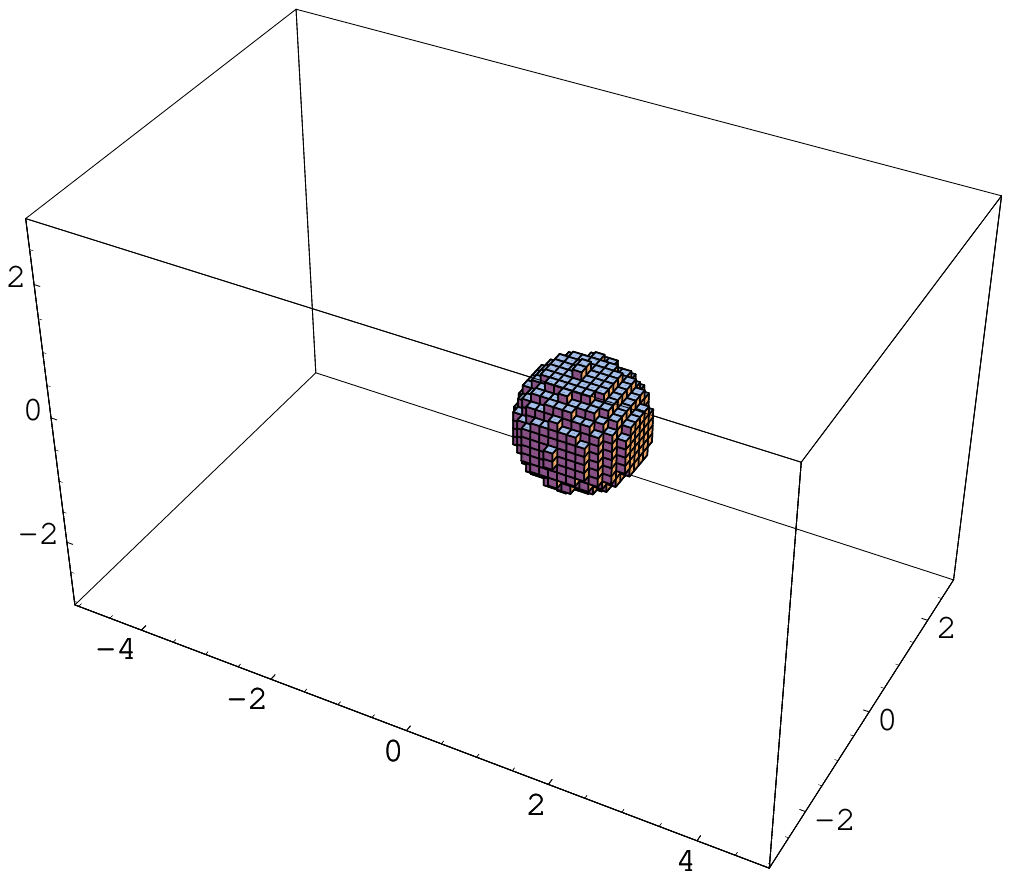}\hfill
\includegraphics[width=0.32\linewidth,bb=0 220 369 470]{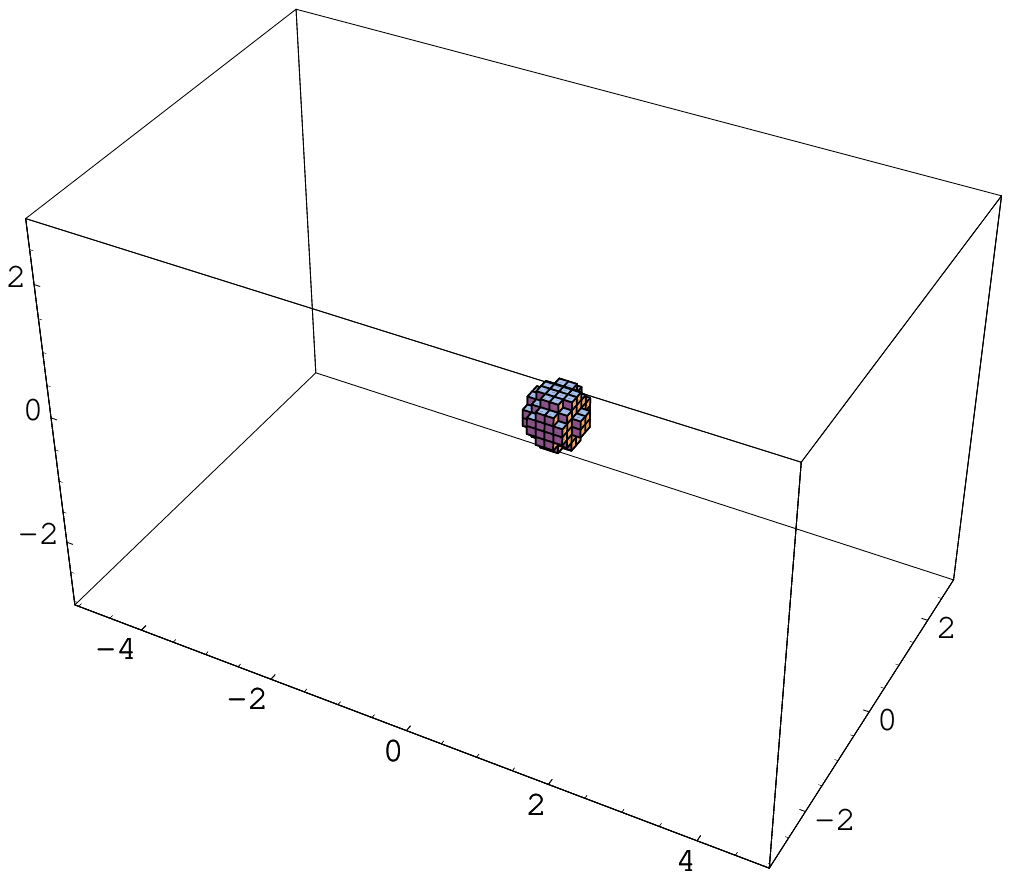}
\includegraphics[width=0.32\linewidth,bb=-60 190 309 440]{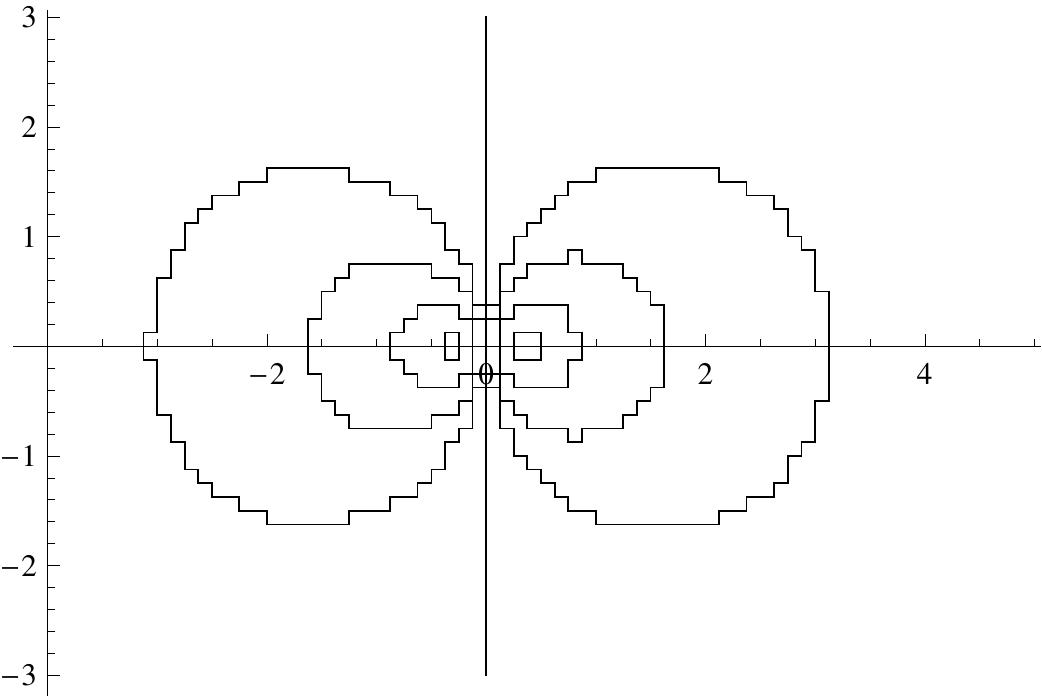}\\
\vspace{1.6cm}
\caption{Twist induced part of the vortices (`bubble') from singlet first excited modes for calorons of size
$\rho=0.6\beta$ in ($x_1$,$x_2$,$x_3$) subspace (a time slice) and plotted in units of $\beta$. Holonomies from left to  right: $\o=0.1,0.12,0.16$ (upper row)
$\o=0.2,0.25,0.3$ (middle row) and $\o=0.34,0.38$ (lower row, left  panel). The plot
in the lower right corner summarizes the results for $\o=0.10,0.12,0.16,0.20,0.25,
0.30,0.34,0.38,0.40$ in a two-dimensional plot at $x_1=0$. The plane near the boundary in the $\o=0.25$ picture
is an artifact caused by periodic boundary conditions.}
\label{twist induced vortices and holonomy}
\end{figure}

The twist induced part of vortices in a unit charge caloron is not much different if
obtained from the doublet or singlet first excited mode. We find that 
the shape of this
part of vortices depends 
%% FB
strongly
%% FB
on $\omega$. For maximal nontrivial holonomy calorons
%% EMI where 
$\omega=1/4$ it is a plane between the two dyons, if $\omega <1/4$ it is a `bubble' 
around the L-dyon, and if $\omega>1/4$ it is a bubble around the M-dyon (all at fixed time $x_0$). The
corresponding pictures are shown in 
Fig.~\ref{twist induced vortices and holonomy}. When this part of vortices is a 
bubble ($\omega \neq 1/4$), the size of the bubble
also depends on $\rho$, see \cite{Caloron and Vortex}.

It might be surprising that the bubble only surrounds one of the dyons while both
dyons are of same importance in a caloron. This is resolved by our claim, that the
bubble is (approximately) the boundary between the static region and the twist
region in a caloron. 

\begin{figure}[t]
\includegraphics[width=0.25\textwidth,bb=-400 150 00 400]{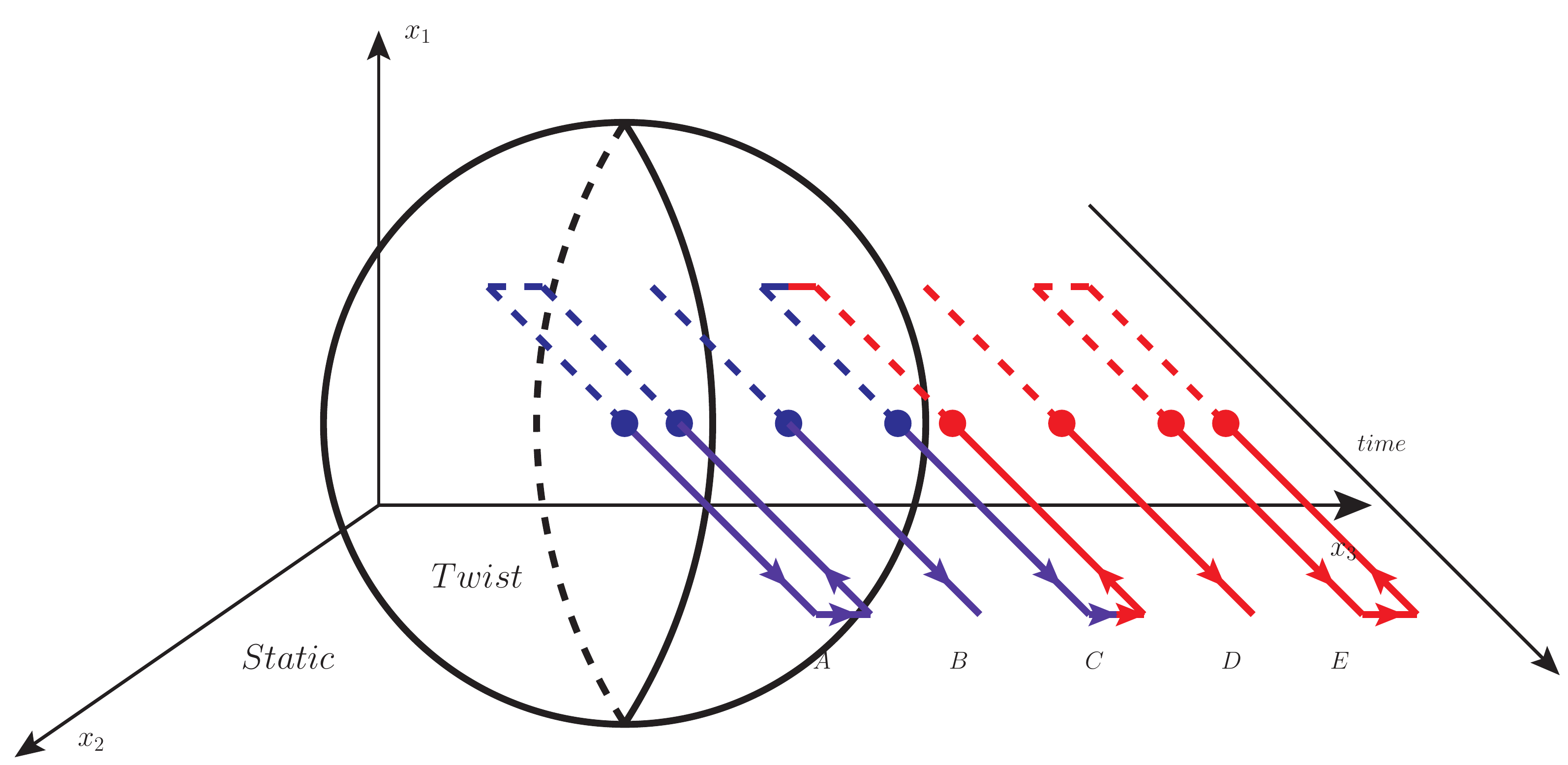}
\vspace{1.0cm}
\caption{Different contours in a pure twisting-static configuration, see text.}
%% FB most part of these contours extend in time direction}
\label{Pure twist static}
\end{figure}

Why is that? Consider a pure twisting-static configuration which mimics the twist:
	\begin{equation}
		A_0(t,\vec{x})=	\left\{
				\begin{array}{ll}
				\pi\sigma_3 / \beta & \,\,\,\,\,	{\rm for} \,\, \vec{x}\in  S\\
				0 &	\,\,\,\,\,  			{\rm for} \,\, \vec{x}\notin  S
				\end{array}
				\right.
		\hspace{1.3cm}
		A_i(t,\vec{x})=0, 
	\end{equation}
where $S\subset \textbf{R}^3$ is the twist region. 
%% FB The signal for the vortices is a
We recall that vortices are signalled by
contractible closed $-\textbf{1}_2$ loops. We 
%% FB can 
consider different contours as 
shown in Fig.~\ref{Pure twist static}. 
%% FB One can find that l
Loops within the static region 
(like D, a Polyakov loop, or E) or contractible loops within twist region (like A) are trivial. 
%% EMI new sentence
On the other hand, 
some loops giving $-\textbf{1}_2$ within twist region (like B, a Polykov loop) are non-contractible.
Only loops that cross the twist {\em and}
%% EMI emphasized "and"
static region (like C) are $-\textbf{1}_2$ and contractible, hence
enclose the vortex surface.
It shows that the vortex surface in this configuration is in space-space direction, 
it is the 2d boundary between different twists, the time coordinate is not fixed by this argument.\\

%Let us consider a thin SU(2) vortices surface intersect with a 2 dimensional surface 
%$\Sigma$ in 4 dimensional space-time. A loop link around the intersection point on
%this 2 dimensional surface should give $-\textbf{1}_2$
%\begin{equation}
%	e^ {\int_{C=\partial \Sigma} i A_\mu dx^\mu}=-\textbf{1}_2
%\end{equation}
%where $\partial \Sigma$ is a closed contour around the intersection point on the 
%2 dimensional surface. 

%\begin{figure}
%\vspace{-2.0cm}
%\includegraphics[width=0.24\linewidth,bb=0 220 369 470]{LCG_CDLk8484880_Om012_Rho060_SpaceSpaceVS_TimeSlice3.pdf}
%\includegraphics[width=0.24\linewidth,bb=0 220 369 470]{LCG_CDLk8484880_Om012_Rho070_SpaceSpaceVS_TimeSlice3.pdf}
%\includegraphics[width=0.24\linewidth,bb=0 220 369 470]{LCG_CDLk8484880_Om012_Rho080_SpaceSpaceVS_TimeSlice3.pdf}
%\includegraphics[width=0.24\linewidth,bb=0 220 369 470]{LCG_CDLk8484880_Om012_Rho090_SpaceSpaceVS_TimeSlice7.pdf}
%\vspace{2.0cm}
%\caption{Spatial part of the vortices (`bubble') for calorons of fixed
%intermediate holonomy $\o=0.12$ and sizes from left to right:
%$\rho=0.6,0.7,0.8,0.9$.}
%\label{twist induced vortices and rho}
%\end{figure}

\begin{figure}[h]
\includegraphics[width=0.6\linewidth,bb=-200 -140 150 110]{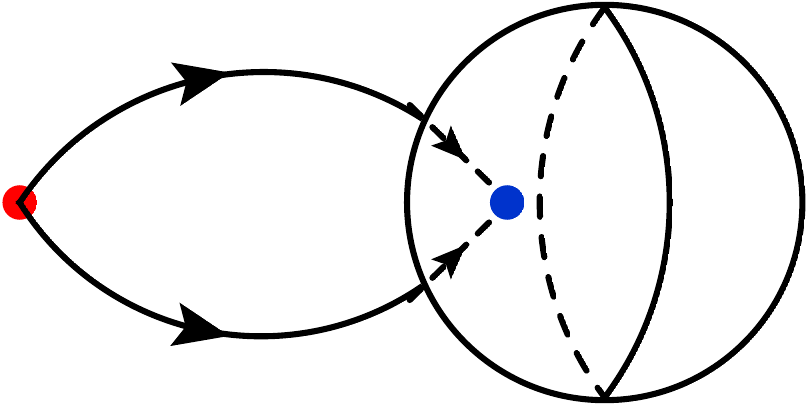}
\vspace{-3.5cm}
\caption{The intersection of the vortices in a caloron at fixed $x_0$ coordinate. Dots indicate the dyons (monopoles in the LAG).}
\label{fig:intersection_schematically}
\end{figure} 

%% FB In the case of large $\rho$, 
%% EMI	The dyon charge induced vortices in large $\rho$ cases is completely 
	The dyon charge induced vortex surface 
%% FB
in the case of large $\rho$
%% FB
%% EMI	in large $\rho$ cases 
is completely 
in space-time direction, while the twist induced vortices is in space-space 
direction. 
%% EMI These two parts of vortices always have two intersection points as shown 
These two kinds of vortices always have two intersection points as shown 
in Fig.~\ref{fig:intersection_schematically}.
Each of these points gives
topological charge\footnote{As well known, an intersection point 
can contribute
%% EMI give 
topological charge $\pm 1/2$.} $1/2$, and this recovers the topological 
charge of the caloron. The general relation of 
vortices with topological charge has been 
worked out 
in \cite{Engelhardt:2000wc}.
%% FB for vortices consisting of hypercubes in \cite{Engelhardt:2000wc} 
%% FB and for smooth vortices in \cite{Engelhardt:1999xw}.

\section{Vortex Surface in Caloron Gases}

The caloron ensembles considered in this paper have been created along the lines
of Ref.~\cite{Gerhold:2006bh}. The 4 dimensional center of mass locations of 
the calorons 
%% FB
and anticalorons 
%% FB
are sampled randomly as well as the spatial orientation of the 
`dipole axis' connecting the two dyons and the angle of a global $U(1)$ 
rotation around the axis $\omega \sigma^3$ in color space. The caloron size is 
sampled from a suitable size distribution $D(\rho,T)$.

\begin{figure}
\includegraphics[width=0.35\textwidth,bb=0 310 369 700]{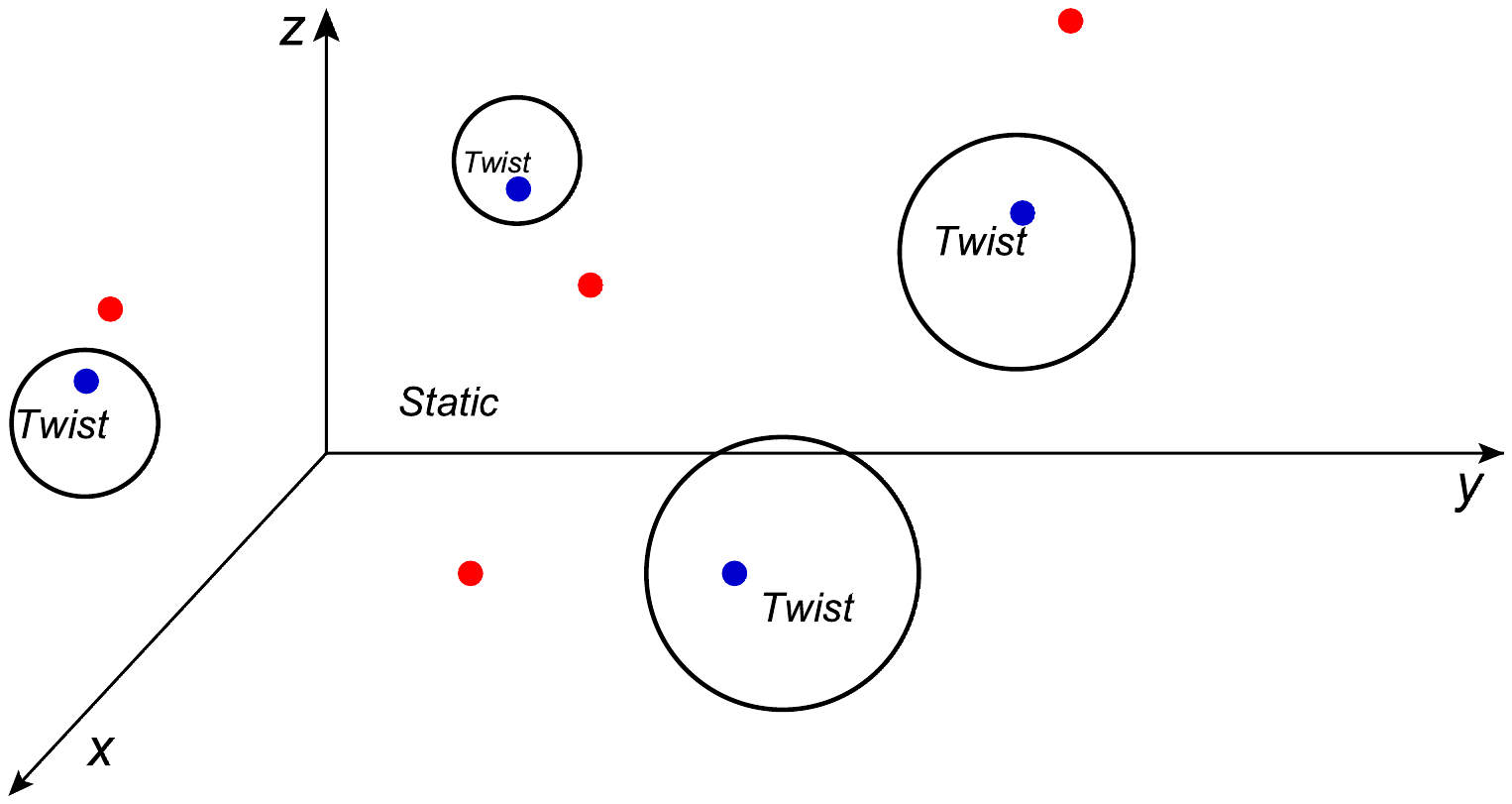}
\includegraphics[width=0.35\textwidth,bb=-100 310 269 700]{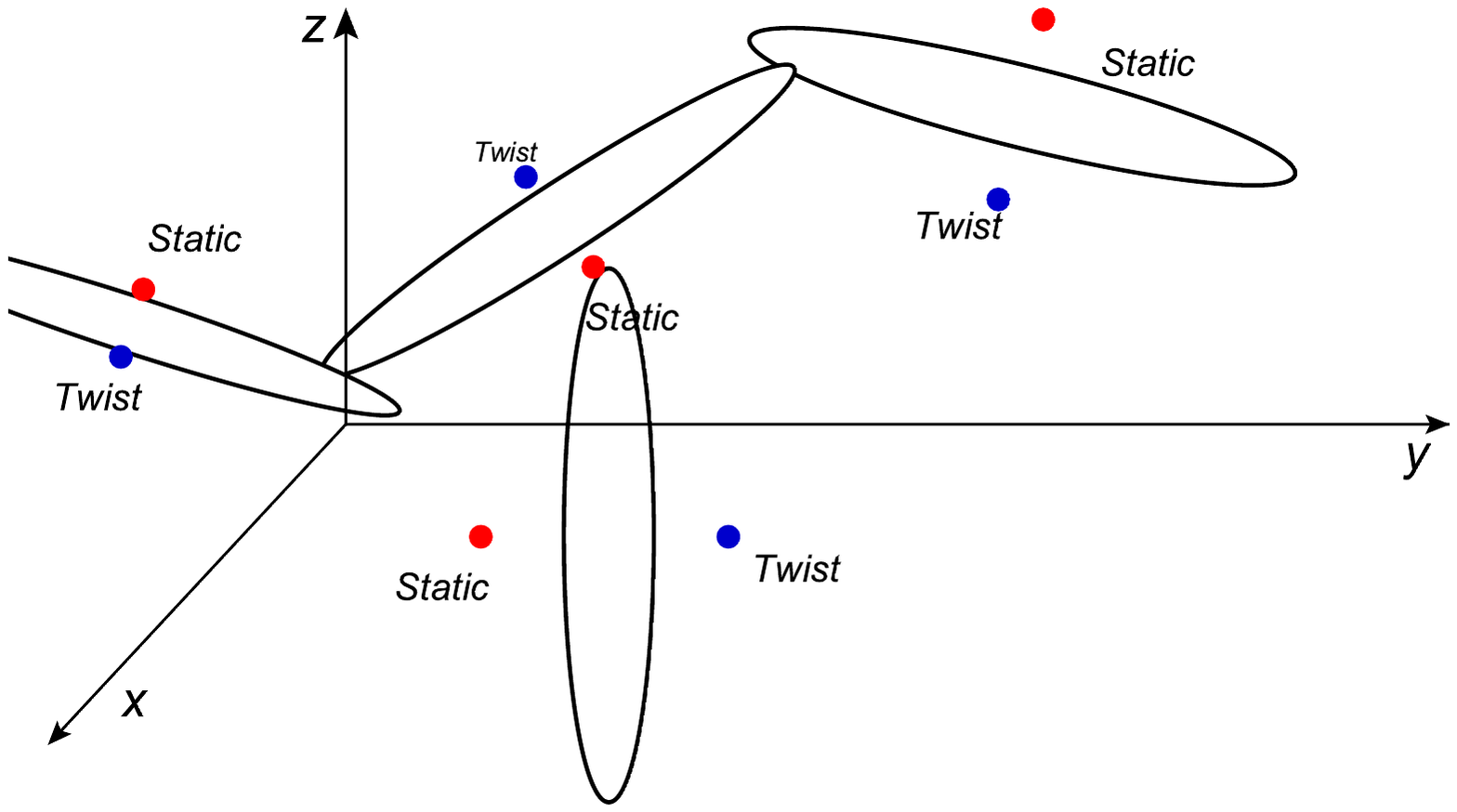}
\vspace{-2.3cm}
\caption{Recombination of vortices in caloron gases (schematically) for holonomy parameters $\omega\neq 1/4$ (left) and maximal nontrivial holonomy $\omega=1/4$  (right).}
\label{recombination of vortices}
\end{figure}

Suppose the vortex distribution in a dilute caloron 
gas forms by a recombination of the vortices of single calorons, see Fig.\ref{recombination of vortices}.
%% FB , we will find that for an $\omega \neq 0.25$ caloron gas,
Then for a caloron gas with holonomy $\omega\neq 1/4$ 
the space-space directed vortices should be separated
bubbles, 
%% FB each 
a bubble for 
%% FB a 
each caloron (if these bubbles are not too 
close or too large
%% EMI enough 
to touch each other). But the 
%% FB case 
situation in a caloron gas
with maximal nontrivial holonomy $\omega = 1/4$
is completely different. The space-space directed
vortices in every single caloron form a plane that 
%% EMI try 
tends to extend to infinity. In the result,
%% EMI is that 
the vortices of different calorons 
%% EMI try to 
touch each other and recombine to
%% EMI thus always
form a percolating surface. 
%% EMI This result 
This scenario is 
%% FB
rather 
%% FB
independent of the
caloron density or temperature.

We recall that vortices give a linear potential only when they percolate. Percolation means the sizes of some of the vortices clusters are comparable to the size of the lattice.
Several papers have reported percolation of 
%% EMI vortices clusters 
vortex clusters in Monte Carlo simulated 
configurations below the critical temperature \cite{percolation}.

\begin{figure}
	\includegraphics[width=0.24\textwidth,bb=60 -20 369 290]{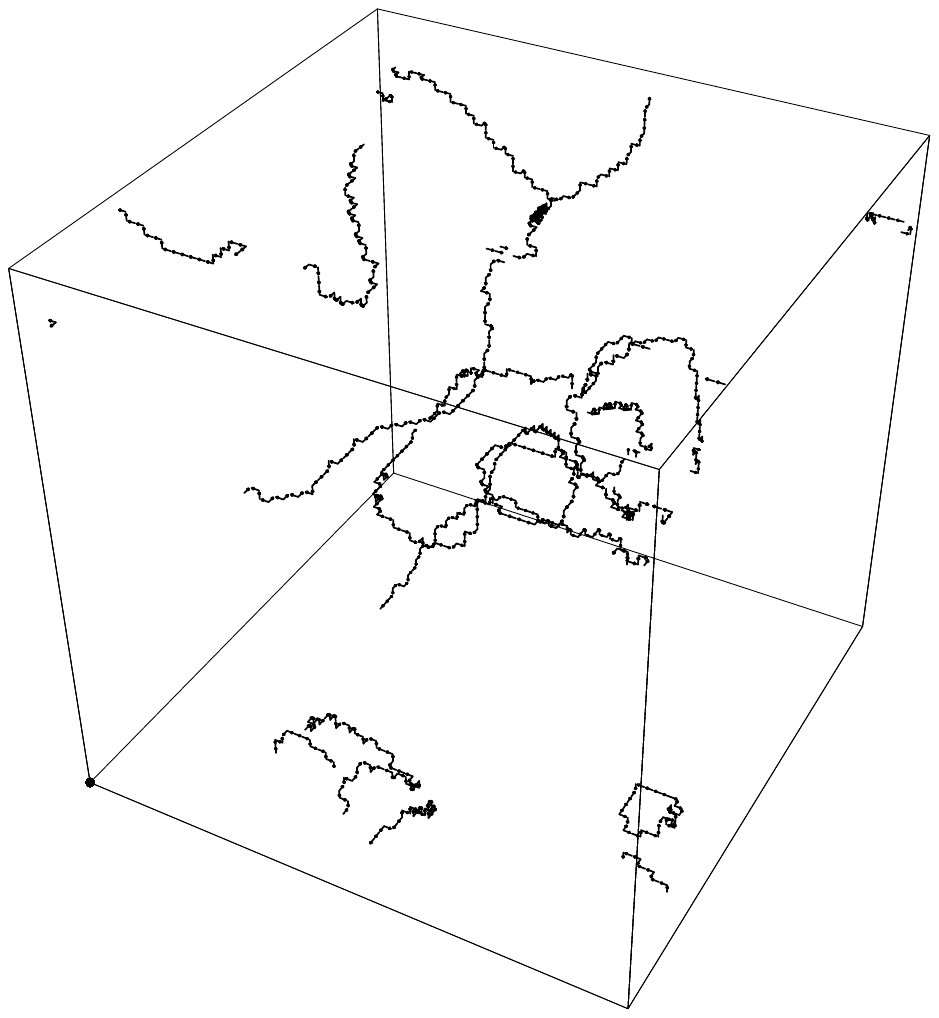}
	\includegraphics[width=0.24\textwidth,bb=60 -20 369 290]{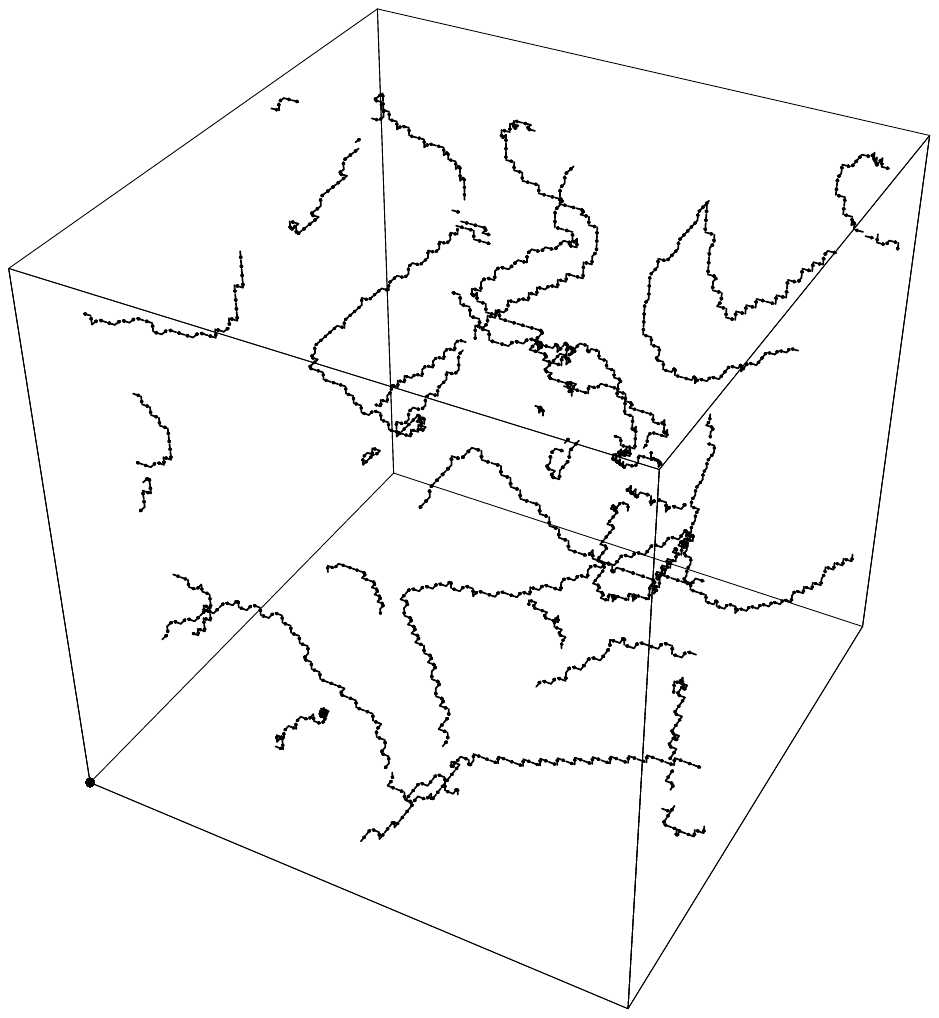}
	\includegraphics[width=0.24\textwidth,bb=60 -20 369 290]{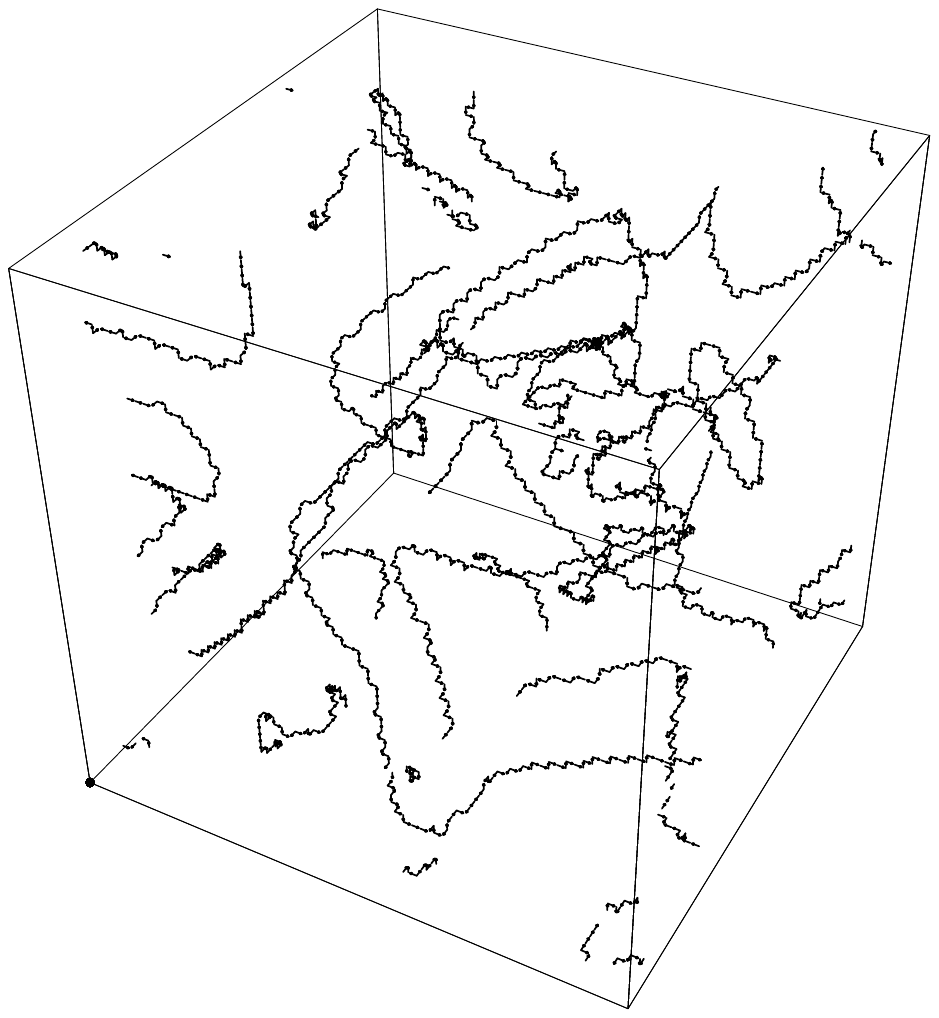}
	\includegraphics[width=0.24\textwidth,bb=60 -20 369 290]{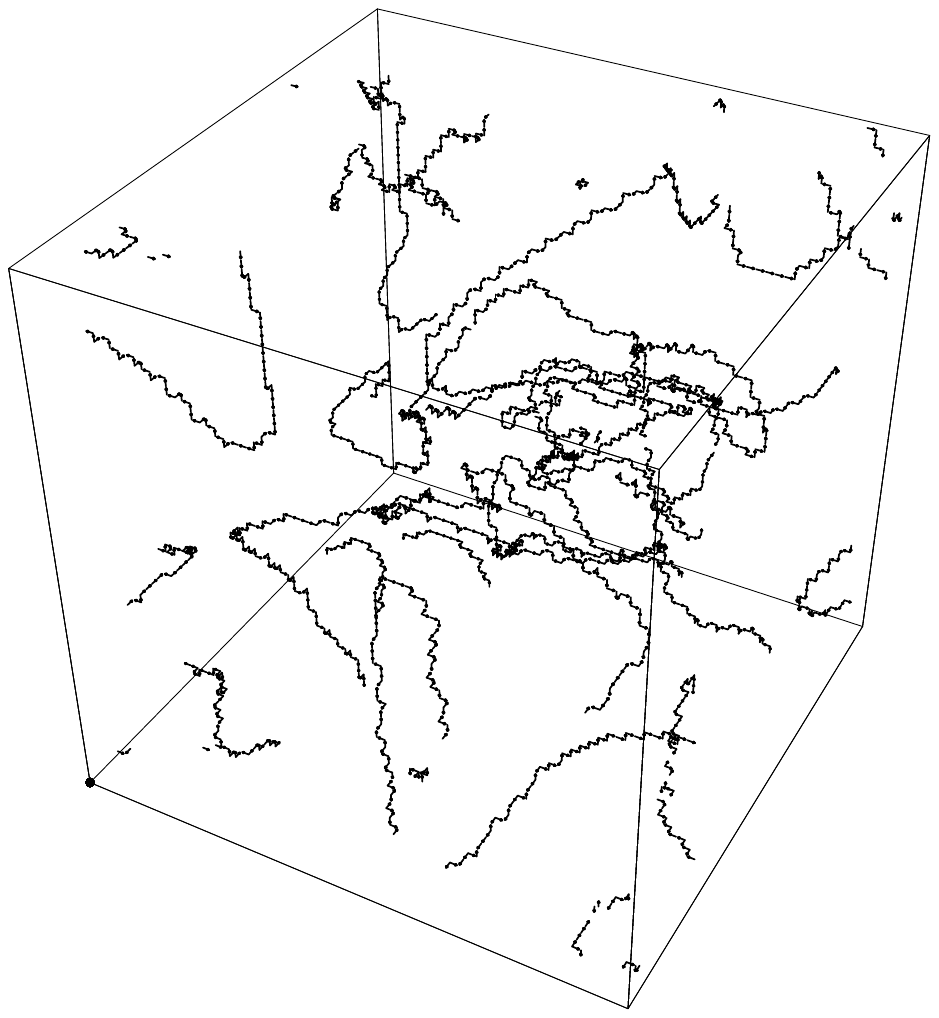}\\
	\includegraphics[width=0.24\textwidth,bb=60 -20 369 290]{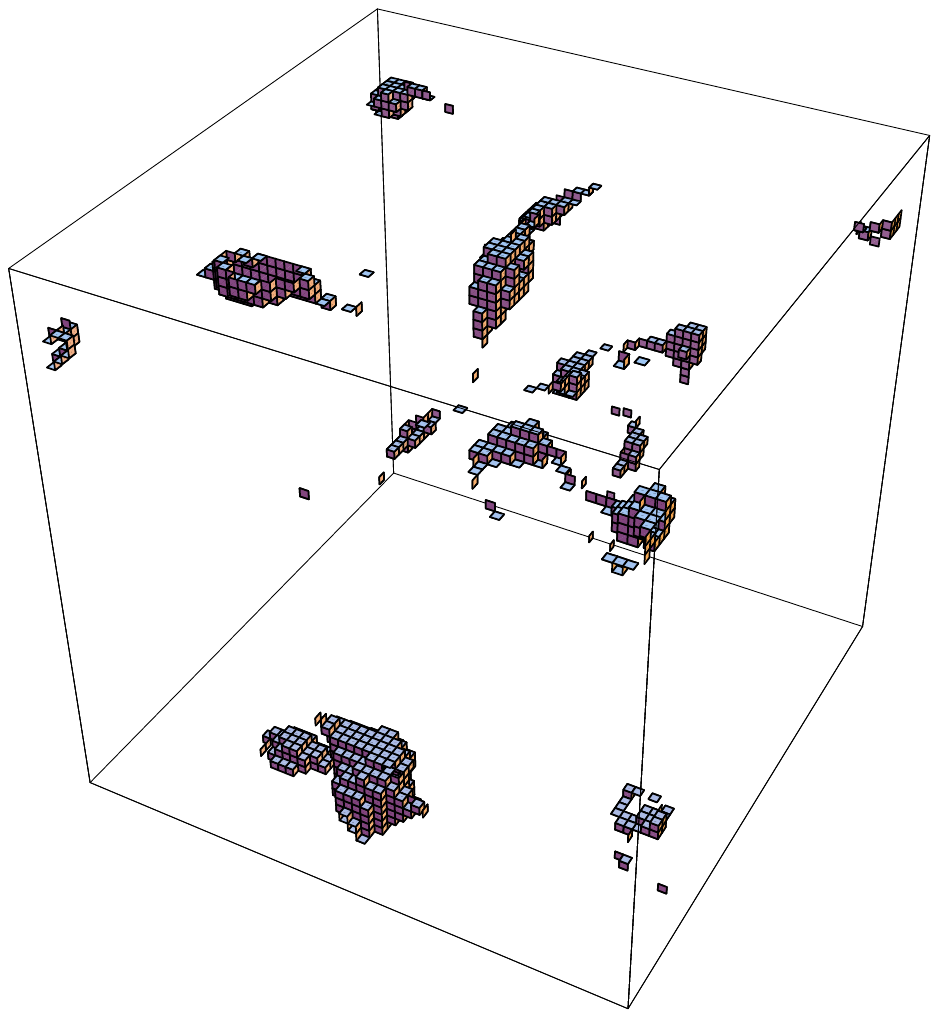}
	\includegraphics[width=0.24\textwidth,bb=60 -20 369 290]{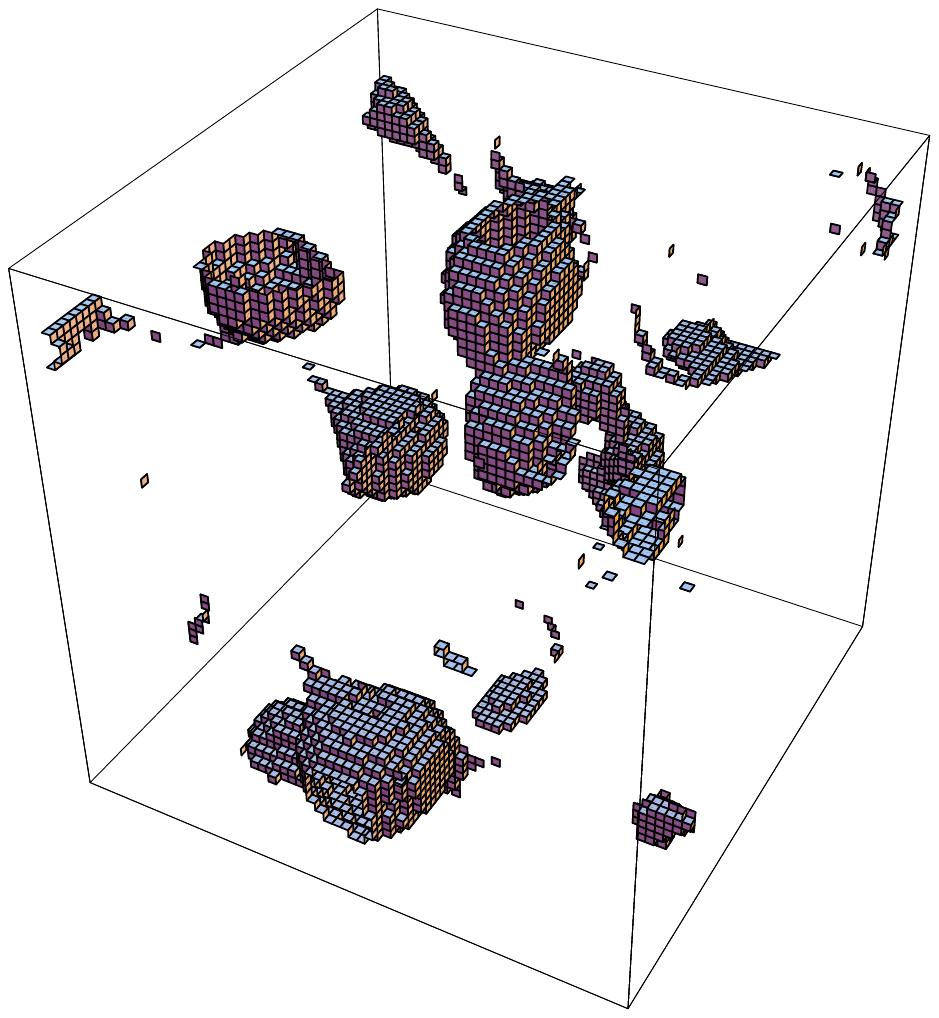}
	\includegraphics[width=0.24\textwidth,bb=60 -20 369 290]{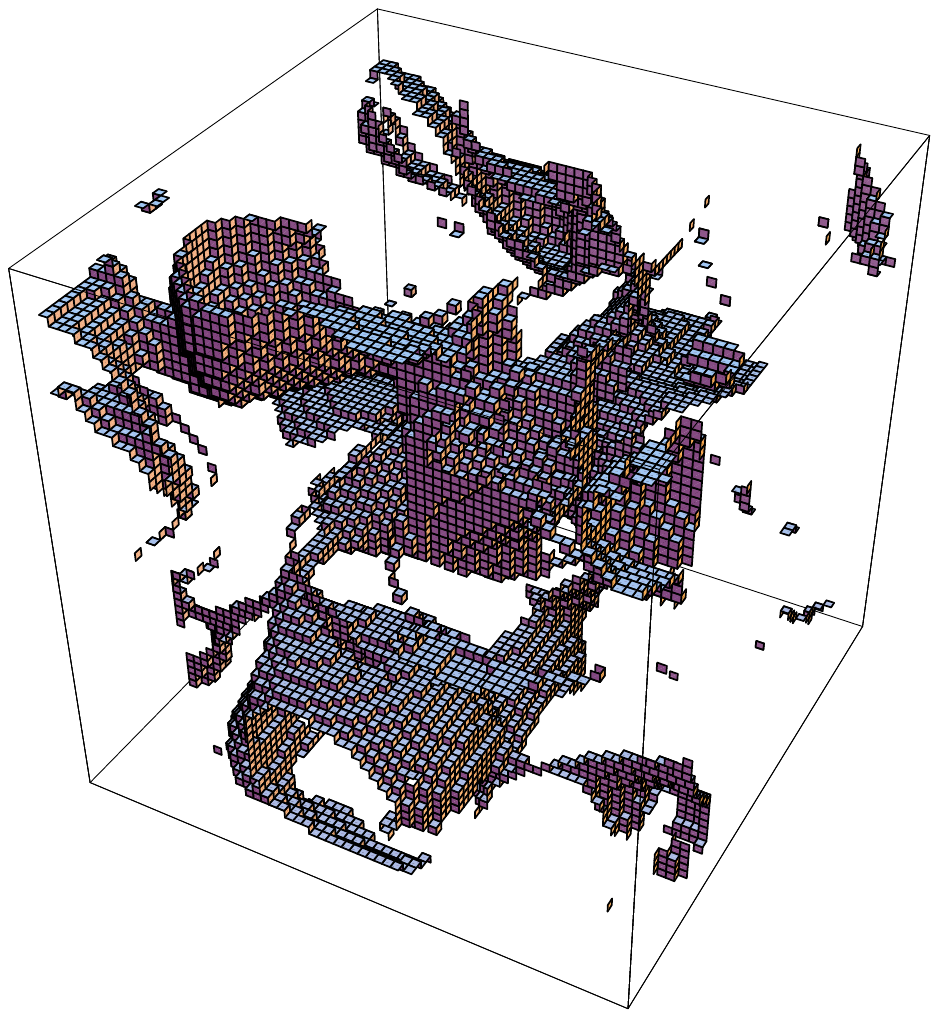}
	\includegraphics[width=0.24\textwidth,bb=60 -20 369 290]{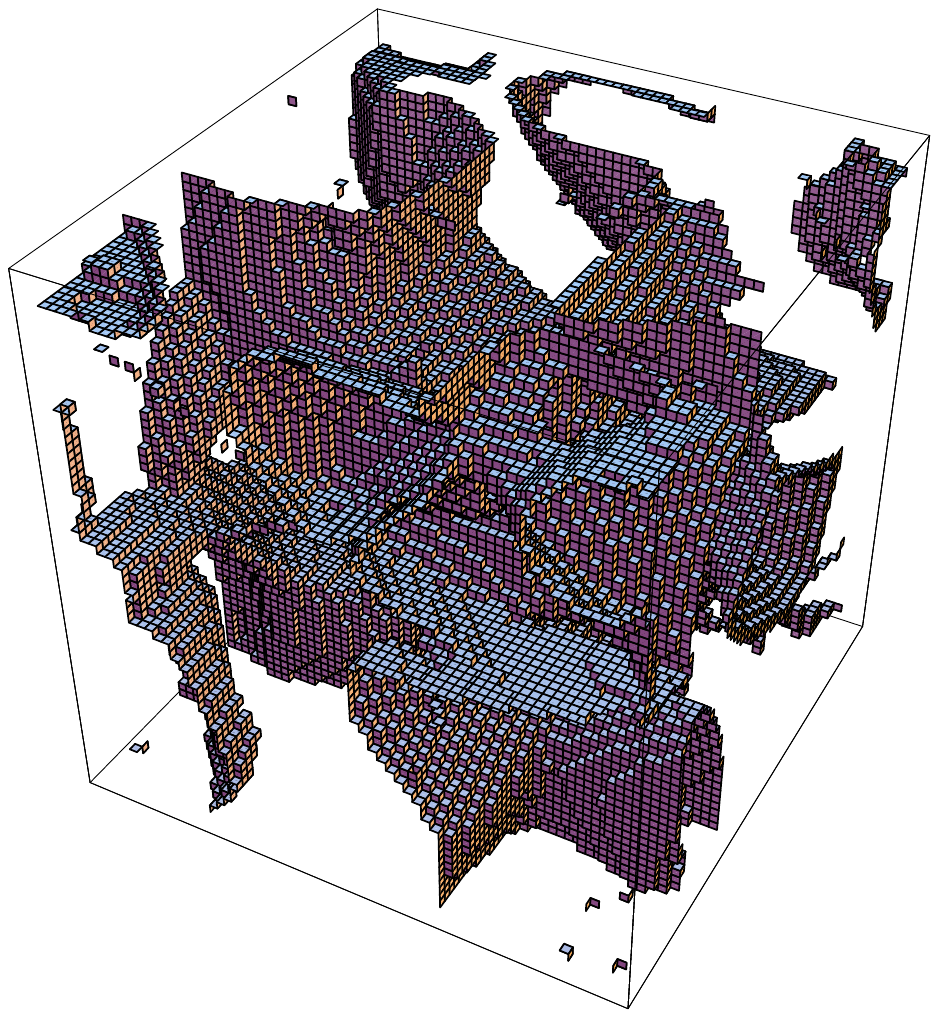}\\
\vspace{-0.6cm}
\caption{Vortices in caloron gas configurations.
The first row are space-time direction vortices traced out within the box representing the first time slice.
The second row are space-space direction vortices from all time slices overlaid in one spatial box.
The holonomy parameter $\omega$ of the (anti)calorons is 0.0625, 0.125, 0.1875 and 0.25 from left to right.
\label{CalGas VS Pictures}}
\end{figure}

Numerical results for vortices in caloron emsembles with different
holonomy are shown in Fig. \ref{CalGas VS Pictures}. These configurations have
their calorons located at the same random coordinates, in the same direction
and with the same $\rho$ parameters,
the only varying parameter is the holonomy $\omega$.
It can be seen clearly that space-space directed vortices only in the maximally nontrivial holonomy case
%ZB ($\omega=1/4$)
form percolating clusters, while the space-time directed
vortices percolate independently of $\omega$. Thus calorons induce confinement for $\omega=1/4$
and a constant string tension of spatial Wilson loops.
Our results via center vortices agree with other caloron studies showing a similar
holonomy-dependence of confinement manifesting itself in Polyakov correlators \cite{gerhold:dublin,Gerhold:2006bh}.

%% FB Numerical results for vortices in caloron gas configurations with different 
%% FB holonomy are shown in Fig. \ref{CalGas VS Pictures}. These caloron gas 
%% FB configurations have 
%% FB their calorons located at the same random coordinates, in the same direction 
%% FB and have the same $\rho$ parameters. 
%% FB %% EMI The only different parameter is the holonomy.
%% FB The only varying parameter is the holonomy $\omega$.
%% FB It is very clear that space-space directed vortices only in a gas 
%% FB %% FB of maximally nontrivial calorons 
%% FB of calorons with maximally nontrivial holonomy $\omega=1/4$
%% FB form percolating clusters, while the space-time directed 
%% FB %% FB ones
%% FB vortices 
%% FB {\tt penetrate} in time direction independent of $\omega$. This result can 
%% FB explain why caloron simulations show 
%% FB %% FB normal 
%% FB confinement for $\omega=1/4$
%% FB and a constant string tension of spatial Wilson loops.
%% FB %% EMI means there is a constant string tension
%% FB %% EMI in maximal nontrivial holonomy caloron gas.
%% FB {\tt do you have some other caloron findings in mind with this sentence?!}

\section{Conclusion}

Using Laplacian Center Gauge, we find that the vortices in a $SU(2)$ caloron include
two parts: the constituent dyon charge induced part and the twist induced part. The latter 
part is mainly in space-space direction and percolates in a caloron ensemble in the case of
maximal nontrivial holonomy. 
Under our conjecture this amounts to the confined phase with vanishing Polyakov loop.
We have demonstrated that this behavior can be understood from
the dependence of the vortex shape on the holonomy in individual calorons.
This finding fits perfectly in the confinement scenarios of vortices and shows
that calorons are suitable to facilitate the vortex confinement mechanism.

This work has been supported by DFG (BR 2872/4-1).

\end{document}